\documentclass[%
reprint,
showpacs,preprintnumbers,
 amsmath,amssymb,
 aps,
pre,
]{revtex4-1}

\usepackage{graphicx}
\usepackage{dcolumn}
\usepackage{bm}

\begin{document}


\title{Energy exchange and localization of low-frequency oscillations in single-walled carbon nanotubes}

\author{V.V. Smirnov$^1$}%
\email{vvs@polymer.chph.ras.ru}
\author{L.I. Manevitch$^1$}
\author{M.Strozzi$^2$}
\author{F.Pelicano$^2$}
 \affiliation{%
 $^1$Institute of Chemical Physics, RAS,
4 Kosygin str., 119991 Moscow, Russia
$^2$ Department of Engineering "Enzo Ferrari", University of Modena and Reggio Emilia, Strada Vignolese 905, 41125 Modena, Italy}%

\date{\today}

\begin{abstract}
We present the results of analytical study and Molecular Dynamics simulation of low energy nonlinear non-stationary dynamics of single-walled carbon nanotubes (CNTs). New phenomena of intensive energy exchange between different parts of CNT and weak energy localization in a part CNT are analytically predicted in the framework of the continuum shell theory. These phenomena take place for CNTs of finite length with medium aspect ratio under different boundary conditions. Their origin is clarified by means of the concept of Limiting Phase Trajectory, and the analytical results are confirmed by the MD simulation of simply supported CNTs.
\end{abstract}

\pacs{61.48.De, 63.22.Gh, 63.20.D-,05.45.-a}
\maketitle



\section{Introduction}\label{Int}

From a modern point of view, carbon nanotubes are attractive subjects for two reasons. 
In the first place, they are associated with great hopes for the creation of super-small and ultra-fast electronic and electromechanical devices with unique physical properties \citep{CLi03, Sazonova2004, Peng2006,  Anantram06}. 
On the other hand, they are quasi-one-dimensional objects, that allows to  check out some of the fundamentals of the modern solid-state physics.
 In particular, variuos computational and in-situ measurements of thermoconductivity of CNT \citep{Berber00, Wang06, Mingo05,BLi05, Savin09} are directly related with the problem of finiteness of thermoconductivity of one-dimensional anharmonic lattices. 
 This problem has been formulated more than fifty years ago by Fermi, Pasta and Ulam \citep{FPU} and it is not completely resolved till now. 
The stationary dynamics of CNTs or nonstationary, but non-resonance dynamics of CNT, can be treated in terms of linear or nonlinear normal modes. 
Using their combinations, one can describe the CNT oscillations under arbitrary initial conditions.
However, the situation drastically changes if we deal with non-stationary resonance processes such as energy transfer. 
In the framework of the linear theory, the energy transfer requires the formation of a wave packet, the time evolution of which depends  strongly on the dispersion properties of the system. 
The dispersion leads to the wave packet spreading that strongly affects  the energy transfer efficiency. 
In the nonlinear systems, the dispersive spreading, can be compensated by nonlinearity. 
As a result, a soliton (breather) mechanism of energy transfer in the infinite quasi-one-dimensional nonlinear lattices arises.
However, it was recently shown \citep{VVS2010, DAN2010, VVS2011} that the resonant interaction of nonlinear normal modes in the finite lattices leads to the  existence of significant non-stationary phenomena, which disappear in the infinite case: i) the intensive energy exchange between different parts of the system, which can be observed at a small enough excitation level, ii) the existence of an instability threshold for the zone-boundering mode, iii) the transition to  weak energy localization in some part of the system. 
All these phenomena can be understood and efficiently described  by a unified viewpoint  in the framework of Limiting Phase Trajectory (LPT) concept. 
The LPTs correspond to strongly non-stationary processes, which are characterized by the maximum possible (under given conditions) energy exchange between different parts of the system. 
In the present paper we show that the processes of the intensive energy exchange and the transition to the energy capture in the some part of the CNT can be explained by the transformation of the LPT with the growth of the excitation level.
The problem of nonlinear resonant interaction of low-frequency vibrational modes of CNT together with the description of phenomena mentioned above are the main subjects of this paper. 
A brief preliminary discussion of the revealed phenomena was presented in \citep{PRL2014}.

\section{The model}
\subsection{Sanders-Koiter thin shell theory and its modification}\label{SKTST}

The dynamics of carbon nanotubes (CNT) is one of the few areas of solid state physics, in which the classical theory of thin elastic shells (TTES) can be legitimately applied. 
It is noteworthy that, in contrast to macroscopic mechanics, where the fundamental limits of TTES are restricted by the possibility of plastic deformation, this theory can also be used for large displacements of CNTs, even in the analysis of their collapse.
The only complicating factor is the uncertainty of the parameter characterizing the thickness of the CNT \citep{Huang06}.
The applicability of a well-designed TTES allows us to obtain an effective description of the vibrational spectrum in the framework of the linear approximation. 
 This can be easily performed  for the simplest  of the boundary conditions, when a CNT of finite size can be considered as a part of an infinite CNT.
  However, the modified theory presented below admits efficient study of both linear and nonlinear dynamics of CNTs under arbitrary boundary conditions.

We have found analytically the corresponding part of the oscillation spectrum and estimated the effect of the boundary conditions taking into account the existence of boundary layers \citep{SoundVibr2014}. 
As for the non-stationary nonlinear problems for CNT, in particular, resonance intermodal interaction, they can be, in principle, studied by numerical methods \cite {Shi08, Shi09, Soltani11}. 
However, such approach is unsufficient when dealing with  prediction of new phenomena and the range of their manifestation. 
On the  contrary, we show that the analytical approach to nonlinear dynamics of the CNT turns out to be efficient from this viewpoint. 

In the considered case an  interaction of lowest (by eigen frequency) CFMs, which is essential due to the effective crowding of the eigenvalues  in the left range of the spectrum, leads to existence of resonance non-stationary process. 
This analytical investigation is based on the reduced nonlinear Sanders-Koiter thin shell theory, and it is a far going extension of our recent study relating to the origin of the oscillation localization in the nonlinear lattices of various types \citep{VVS2010, DAN2010, VVS2011}.

Two optical-type vibration branches in the CNT spectrum (fig. 1) are of interest from the viewpoint of the energy exchange processes mentioned above: the one of them is the well-known Radial Breathing Mode (RBM), which is associated with the circumferential wave number n = 0 and corresponds to uniform radial extension-compression.
The  lowest optical mode in the CNT spectrum (Circumferential Flexure Mode - CFM) [18] is specified by n = 2, and the main deformation is a deviation of the CNT cross-section from the initial circular one \citep{Rao97}.  
To the best of our knowledge, the nonlinear dynamic processes on CNTs were analytically studied only on the basis of a simplest modal analysis (RBM and its parametric instability) \cite {Shi08, Shi09, Soltani11}.

\begin{figure}

\centering{
 \includegraphics[width=70 mm]{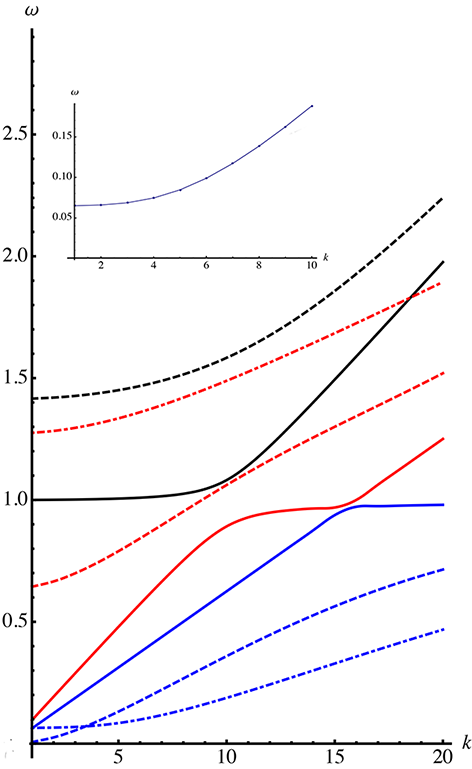}
}
\caption{
The CNT spectrum according to the exact Sanders-Koiter thin shell theory: solid curves correspond to circumferential wave number $n=0$, dashed ones - to $n=1$ and dot-dashed one - to $n=2$. 
The insert shows the small wave number part of the CFM branch.
 All the frequencies $\omega$ are measured in dimensionless units and  $k$ - denotes the number of longitudinal half-waves along the CNT.}
\label{fig:CNT2chain}
\end{figure}

 The CFM oscillations are characterized by relative smallness of the ring and shear deformations  (in particular, the contour length of the lateral section is not essentially changed during deformation). 
 So, we can assume that only the bending, torsion and longitudinal deformation contribute to the potential energy. 
 

The energy of elastic deformation of CNT in the dimensionless units is written as follows:


\begin{equation}\label{energy_elst2}
\begin{split}
E_{el} =\frac{1}{2} \int\limits_{0}^{1} \int\limits_{0}^{2 \pi}\left(\varepsilon_{\xi}^{2} + \varepsilon_{\varphi}^{2} + 2 \nu \varepsilon_{\xi} \varepsilon_{\varphi} + \frac{1 - \nu}{2} \varepsilon_{\xi \varphi}^{2}\right)d\xi d\varphi +\\
+\frac{\beta^{2}}{24} \int\limits_{0}^{1} \int\limits_{0}^{2 \pi}\left(\kappa_{\xi}^{2} + \kappa_{\varphi}^{2} + 2 \nu \kappa_{\xi} \kappa_{\varphi} + \frac{1 - \nu}{2} \kappa_{\xi \varphi}^{2}\right)d\xi d\varphi ,
\end{split}
\end{equation}

where $\varepsilon_{\xi}$, $\varepsilon_{\varphi}$ and $\varepsilon_{\xi \varphi}$ are the longitudinal, circumferential and shear deformations, and $\kappa_{\xi}$, $\kappa_{\varphi}$, $\kappa_{\xi \varphi}$ are the longitudinal and circumferential curvatures, and torsion, respectively.


 Considering the small-amplitude oscillations of the CNT in the limiting case of a large aspect ratios, one can write the following equation of motion in terms of the radial component of the displacement (see Appendix \ref{app:A} for details)
 
\begin{equation}\label{eq:WWW}
\begin{split}
 \frac {\partial^{2}W}{\partial \tau_{0}^{2}}+   W -\varepsilon^{2}\frac{\mu}{\omega_{0}^{2}} \frac{\partial^{2} W}{\partial \xi^{2}}   - \varepsilon^{2} \gamma\frac{\partial ^{4} W}{\partial \xi^{2} \partial \tau_{0}^{2}}  
 +\varepsilon^{2}\frac{\kappa}{\omega_{0}^{2}}  \frac{\partial^{4} W}{\partial \xi^{4}}   \qquad \quad \\
 + a_{1} W \left( \left( \frac{\partial W}{\partial \tau_{0}} \right) ^{2}+W \frac{\partial ^{2} W}{\partial \tau_{0}^{2}} \right)  
 + \varepsilon^{2} \frac{a_{2}}{\omega_{0}^{2}} \left( \frac{\partial W}{\partial \xi} \right)^{2} \frac{\partial^{2} W}{\partial \xi^2} = 0,
 \end{split}
\end{equation}


where $W$ characterizes the radial displacement of the shell,  $\alpha$ is the inverse aspect ratio of the CNT (i.e. the ratio of the CNT radius $R$ to its length $L$), $\beta$ is the ratio of the thickness of the CNT wall $h$ to its radius and $\omega_{0}$  is the gap frequency.
Other parameters depend on the circumferential wave number $n$ and the Poisson ratio $\nu$. 
 $\xi$ and $\tau_{0}=\omega_{0} \tau$ are the dimensionless coordinates along the CNT axes ($0 \le \xi \le 1$) and the dimensionless time reduced to the gap frequency $\omega_{0}$, respectively. 
 One should note that, taking into account the small value of $\omega_{0} \thicksim \beta \ll 1$ and the large aspect ratio $\alpha \ll 1$, we clearly assign the order of smallness of the different terms in equation \eqref{eq:WWW}. 
 At the same time we formally  consider that the coefficients of equation \eqref{eq:WWW} to be of the zero order by small parameter.


Equation \eqref{eq:WWW} is a useful tool to analyze the effect of various boundary conditions on the spectrum of natural oscillations of the CNT \citep{Polymer2013} (see brief discussion in the Appendix \ref{app:B}).
The frequency spectrum in the case of simply supported edges is determined by the following expression:

\begin{equation}\label{eq:omega}
\begin{split}
\omega^{2}=\frac{ \omega_{0}^{2}+\mu\ \pi^{2} k^{2}+ \kappa \pi^{4} k^{4}}{1+ \gamma \pi^{2}  k^{2}},
\end{split}
\end{equation}


where $k$ is a longitudinal wave number corresponding to the number of half-waves along the CNT axis.

It is convenient to rewrite the equation \eqref{eq:WWW} using complex variables:

\begin{multline}\label{eq:complex1}
\Psi=\frac{1}{\sqrt{2}} \left( \frac{\partial W}{\partial \tau_{0}}+ i W \right) \\
W= \frac{-i}{\sqrt{2}}\left( \Psi - \Psi^{*} \right) \qquad
\frac{\partial W}{\partial \tau_{0}}=\frac{1}{\sqrt{2}}\left( \Psi+ \Psi^{*} \right),
\end{multline}

where the asterisk denotes the complex conjugation.

Performing the multiscale expansion procedure (see Appendix \ref{app:A1}) one can get the equation for the amplitude of the main order in the ''slow'' time $\tau_{2}= \varepsilon^{2} \tau_{0}$: 

\begin{equation}\label{eq:NLSE}
i \frac{\partial \chi_{0}}{\partial \tau_{2} } - \frac{\mu-\omega_{0}^{2} \gamma}{2 \omega_{0}^{2}} \frac{\partial ^{2} \chi_{0}}{\partial \xi^{2}} + \frac{\kappa}{2 \omega_{0}^{2}} \frac{\partial^{4} \chi_{0}}{\partial \xi^{4}} - \frac{a_{1}}{2} |\chi_{0}|^{2} \chi_{0} =0,
\end{equation}

where the main order value $\chi_{0}$ is coupled with the complex function $\Psi = \varepsilon \chi_{0} \exp{(-i \tau_{0})}$, and the small parameter $\varepsilon \thicksim \alpha$.

First of all,  equation \eqref{eq:NLSE} admits the plane-wave solution $\chi_{0}= A \exp{(-i \left( \omega \tau_{2} - k \xi) \right)}$ with the dispersion ratio

\begin{equation}\label{eq:NLSE_dispersion}
\omega= \frac{(\mu-\omega_{0}^{2} \gamma) k^{2}+\kappa k^{4}}{2 \omega_{0}^{2}} - \frac{a_{1}}{2} A^{2},
\end{equation}

where $A$ is the amplitude. 
As it can be seen, this dispersion relation is in accordance with the relation \eqref{eq:omega}. 

Equation \eqref{eq:NLSE} is the modified Nonlinear Schr{\"o}dinger Equation (NLSE), in the the standard version of which the fourth derivative is absent.
As it is well known, the standard NLSE admits the localized solution - the envelope soliton or the breather. 
The presence of fourth derivative complicates the problem, but using Pade approximation \citep{Pade} one can obtain the following localized solution

\begin{equation}\label{eq:NLSEsolution}
\chi_{0}=X_{0} e^{-i \omega \tau_{2}}  \text{sech}(\lambda \xi ) ,
\end{equation}

where

\begin{equation}\label{eq:NLSEparam}
\begin{split}
\lambda = \sqrt{\frac{\mu-\gamma  \omega _0^{2}-\sqrt{\left( \mu -\gamma  \omega _0^{2} \right)^{2}+8 \kappa \omega  \omega _0^{2} } }{2 \kappa }}  \\
X_{0}=\frac{2 \sqrt{2}}{\omega _0 }  \sqrt{\frac{-9 \lambda ^2 \left(\mu -\gamma  \omega _0^2\right)-20 \omega _0^2 \omega}{a_{1}} } 
\end{split}
\end{equation}

Solution \eqref{eq:NLSEsolution} describes a set of soliton-like excitations, which are parametrized by the ''frequency'' parameter  $\omega$.
The permissible values of  $\omega$ are determined by the conditions of the reality of the magnitude  ($1/ \lambda$) (the soliton width) as well as of the amplitude $X_{0}$.
Therefore, these values have to be negative.
It is a natural requirement because the localized solutions can exist in the gap of the vibration spectrum.

\begin{figure}
(a) \includegraphics[width=60mm]{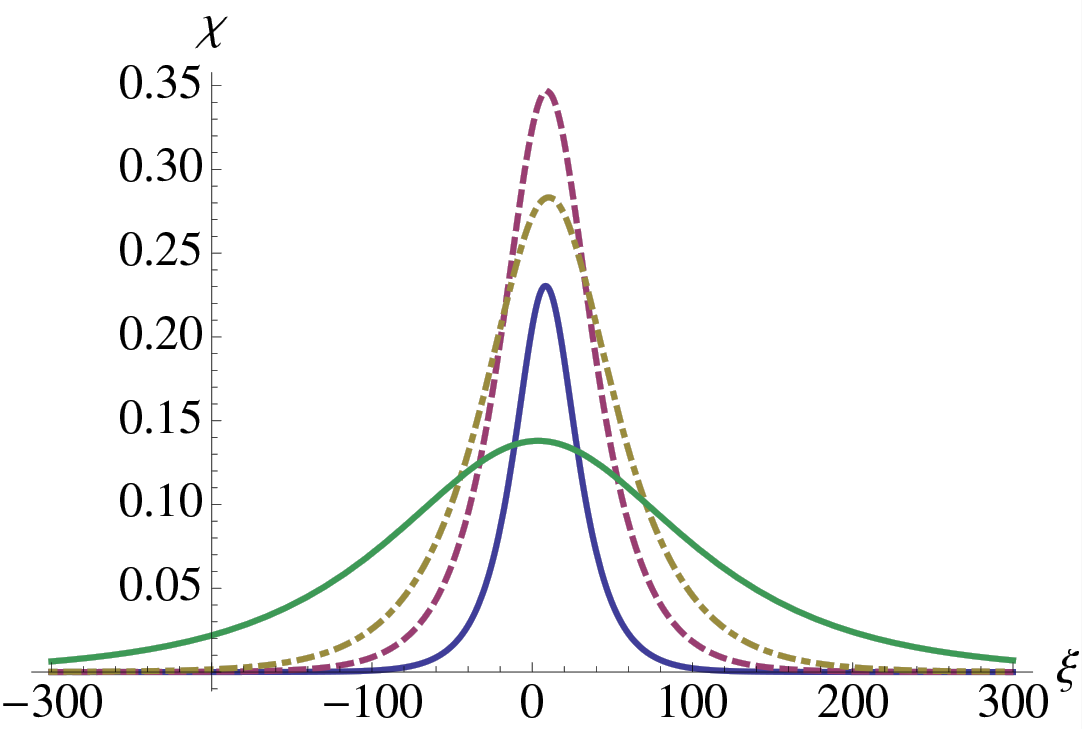} \\
(b) \includegraphics[width=60mm]{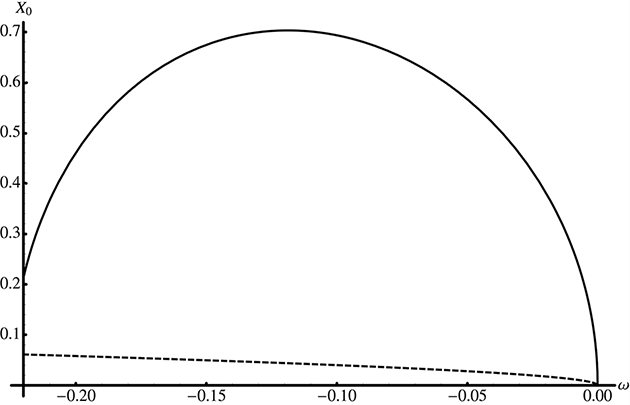}
\caption{ (Color online) (a) The ''soliton'' solution of Eq. \eqref{eq:NLSE} at the variuos values of the ''frequency'' $\omega$ ($-0.2, -0.1, -0.05, -0.01$); (b) The amplitude (thick curve) and the inverse width (dashed curve) of the solution \eqref{eq:NLSEsolution} vs the ''frequency'' $\omega$. }    
\label{fig:soliton}
\end{figure}

Figures (\ref{fig:soliton}) show the solution \eqref{eq:NLSEsolution} and its parameters $\lambda$ and $X_{0}$ at the various values of the frequency $\omega$.
Equation \eqref{eq:NLSE} describes the nonlinear dynamics in the asymptotic limit of the infinitely long CNT. 
Commonly speaking, it is the only limit when the localized soliton-like excitations occur in the nonlinear one-dimensional systems, keeping in mind the boundary conditions in the infinity. 
The modification of equation \eqref{eq:NLSE} is needed to compare the analytical results and the data of numerical or physical experiments. 
One should note that the source of the nonlinearity in equation \eqref{eq:NLSE} is the inertial part of the energy. 
However, the finiteness of the CNT requires taking into account the nonlinear part of the elastic deformation energy also. 
As it is mentioned in the Appendix \ref{app:A} the most essential contribution arises from the nonlinear term $\left( \partial W/ \partial \xi \right)^{2} \partial ^{2} W / \partial \xi^{2} $.
Taking into account this contribution one can modify the equation \eqref{eq:NLSE}:

\begin{multline}\label{eq:NLSE_mod}
i \frac{\partial \chi_{0}}{\partial \tau_{2} } - \frac{\mu-\omega_{0}^{2} \gamma}{2 \omega_{0}^{2}} \frac{\partial ^{2} \chi_{0}}{\partial \xi^{2}} + \frac{\kappa}{2 \omega_{0}^{2}} \frac{\partial^{4} \chi_{0}}{\partial \xi^{4}}   \\  
-  \frac{a_{1}}{2} \left| \chi_{0} \right| ^{2} \chi_{0} +\frac{a_{2}}{4} \frac{\partial }{\partial \xi} \left( \left| \frac{\partial \chi_{0}}{\partial \xi} \right|^{2} \frac{ \partial \chi_{0}}{\partial \xi} \right) = 0,
\end{multline}

Equation \eqref{eq:NLSE_mod} like  equation \eqref{eq:NLSE}  admits the plane-wave as well as the soliton-like solution \eqref{eq:NLSEsolution}.
Only the effective amplitude of soliton $X_{0}$ is modified as:

\begin{equation}\label{eq:NLSEmod_sol}
X_{0}=\frac{4}{\omega _0 }  \sqrt{\frac{\kappa \left( 9 \lambda ^2 \left(\mu -\gamma  \omega _0^2\right)+20 \omega _0^2 \omega \right)}{3 a_{2} \lambda ^2 \left(\mu -\gamma  \omega _{0}^{2} \right)-2 \left(a_{1} \kappa -3 a_{2} \omega _{0}^{2} \omega \right) } }
\end{equation}
 
 One should note that the low-frequency limit $\omega \rightarrow 0$ leads to the same values for the parameters of soliton-like solutions both for equation \eqref{eq:NLSE} and for equation \eqref{eq:NLSE_mod}:
 
 \begin{equation}\label{eq:NLSE_lowlimit}
 \begin{split}
 \lambda = \omega_{0} \sqrt{\frac{2 \omega}{ \gamma \omega_{0}^{2} - \mu}} \\
 X_{0} = 4 \sqrt{\frac{-\omega}{ a_{1}}}
 \end{split}
 \end{equation}
 
Nonlinear equation \eqref{eq:NLSE_mod} can be used for the analysis of nonlinear normal modes interaction and, in particular, for finding the transition between two regimes - the intensive energy exchange and energy localization \citep{PRL2014}. 
To perform this, one should take into account that the vibration spectrum for any CNT with a finite length is discrete, i.e. the longitudinal wave numbers are integers.  
To consider the intermodal interaction let us use the sum of the plane waves with the wave numbers $k_{1}$ and $k_{2}$. 

\begin{equation}\label{eq:NLSE2waves}
\chi_{0}=\chi_{01}(\tau_{2}) \sin{(\pi k_{1} \xi)}+\chi_{02}(\tau_{2}) \sin{(\pi k_{2} \xi)}
\end{equation}


Substituting solution \eqref{eq:NLSE2waves} into equation \eqref{eq:NLSE_mod} one should use the Galerkin procedure to obtain the equations for complex amplitudes $\chi_{01}$ and $\chi_{02}$:
\begin{equation}\label{eq:2waves_1}
\begin{split}
i \frac{\partial \chi_{01}}{\partial \tau_{2}} + \delta \omega_{1} \chi_{01} -\frac{3 \sigma_{11}}{16}  \left|\chi_{01} \right|^{2} \chi _{01}  \qquad \qquad \qquad \quad  \\  
-\frac{\sigma_{12}}{8}  \left( 2\left| \chi_{02} \right| ^{2} \chi _{01} + \chi_{02}^{2} \chi _{01}^{*} \right) = 0 \\
i \frac{\partial \chi_{02}}{\partial \tau_{2}} + \delta \omega_{2} \chi_{02} - \frac{3  \sigma_{22} }{16} \left|\chi_{02} \right|^{2} \chi _{02}  \qquad \qquad \qquad \quad  \\
-\frac{\sigma_{12}}{8} \left( 2 \left|\chi_{01} \right|^{2} \chi _{02} +  \chi_{01}^{2} \chi _{02}^{*} \right) = 0,
 \end{split}
 \end{equation}
 where 
 $$
 \delta \omega_{i} = \frac{\mu-\omega_{0}^{2} \gamma}{2 \omega_{0}^{2}} \pi^{2} k_{i}^{2}+\frac{\kappa}{2 \omega_{0}^{2}} \pi^{4} k_{i}^{4} , \quad i=1,2 
 $$

 are the  intervals between the mode frequency and the boundary frequency $\omega_{0}$ of the considered brunch  and
 
 \begin{equation*}\label{eq:sigma}
 \begin{split}
 \sigma_{ij} = \left(2 a_{1}+\pi ^4 a_2 k_{i}^{2} k_{j}^{2} \right)  \quad (i,j=1,2).
 \end{split}
 \end{equation*}
 
 (One can estimate that the frequency shift between the lowest modes ($k_{1}=1, k_{2}=2$) is approximately twice smaller than that for the next pair of modes ($k_{2}=2, k_{3}=3$).)

It is easy to see that the nonlinear terms in equations \eqref{eq:2waves_1} are separated into two groups: the terms $|\chi_{0j}|^{2} \chi_{0i}$  $(i,j = 1,2)$ determine the nonlinear frequency shift, while the terms  $\chi_{0i}^{2} \chi_{0j}^{*} ( i \ne j) $ describe the nonlinear interaction between modes. 
The Hamiltonian corresponding to equations \eqref{eq:2waves_1} can be written as 

\begin{equation}\label{eq:NLSEhamiltonian2}
\begin{split}
H= \delta \omega _1\left|\chi _{01} \right|^2+\delta \omega _2\left| \chi _{02} \right|^2  - \frac{3}{32} \left( \sigma_{11} \left| \chi _{01} \right|^4 + \sigma_{22} \left| \chi _{02} \right|^{4} \right)  \\
-\frac{\sigma_{12}}{16} \left( 4 \left| \chi _{01} \right|^2 \left| \chi _{02} \right|^{2}+ \left(\chi _{02}^{*2} \chi _{01}^2+\chi _{01}^{*2} \chi _{02}^2 \right) \right) \\
\end{split}
\end{equation}

Equations \eqref{eq:2waves_1}, besides the obvious energy integral \eqref{eq:NLSEhamiltonian2}, possess another integral 

\begin{equation}\label{Occupation}
X=\left| \chi_{01} \right|^{2}+ \left| \chi_{02} \right|^{2},
\end{equation}

which characterises the excitation level of the system, and it is an analogue of the occupation number integral in quantum-mechanical terminology.

\subsection{LPT and the localilization of the CNT vibrations}\label{s:LPT}
 As it was shown in \citep{VVS2010, PRL2014}, the modal analysis becomes inadequate at the resonance conditions. 
 Therefore we introduce new variables as the linear combinations of resonating modes with preservation the integral $X$:

\begin{equation}\label{psi}
\phi_{1}=\frac{1}{\sqrt{2}}(\chi_{01}+\chi_{02}); \phi_{2}=\frac{1}{\sqrt{2}}(\chi_{01}-\chi_{02}).
\end{equation}
 
The new variables describe the dynamics of some parts of the CNT \citep{PRL2014} (similary to some groups (clusters) of the particles in the effective discrete one-dimensional chain   \citep{VVS2010, DAN2010, VVS2011}). 
Considering the distribution of energy along the nanotube one can see that such a linear combination of NNMs describes a predominant energy concentration in certain region of the CNT, while the other part of CNT has a lower  energy. 
Because of small difference between frequencies of the modes, the selected  parts of CNT demonstrate a coherent behavior similar to beating in the system of two weakly coupled oscillators. 
Therefore we can consider these regions as new large-scale elementary blocks, which can be identified as unique elements of the system - the ''effective particles'' \citep{VVS2010}. 
The existence of integral of motion \eqref{Occupation} allows to reduce the dimension of the phase space  up to 2 variables - $\theta$ and $\Delta$, which characterize the relationship between the amplitudes and the phase shift   between the effective particles, respectively:

\begin{equation}\label{Angle1}
\phi_{1}=\sqrt{X} \cos{\theta}e^{-i\Delta /2} ;  \quad
\phi_{2}=\sqrt{X} \sin{\theta}e^{  i\Delta /2}.
\end{equation}


Substituting these expressions into equations \eqref{eq:2waves_1}, the equations of motion in the terms of "angular" variables ($\theta, \Delta$) can be obtained:

\begin{widetext}
\begin{equation}\label{eq:NLSEangle}
\begin{split}
\sin{2 \theta } ( \frac{\partial \theta}{\partial \tau_{2}} - \frac{1}{2} ( \delta \omega _{2} -\delta \omega _{1} ) \sin{\Delta}     - \frac{1}{128} X (3 \sigma _{11} ( \sin{2 \Delta } \sin{2 \theta} + 2 \sin{\Delta } )   \\  +3 \sigma _{22} ( \sin{2 \Delta } \sin{2 \theta }-2 \sin{\Delta } )-4 \sigma _{12} \sin{2 \Delta } \sin{2 \theta } ) ) = 0    \\
\sin{2 \theta} \frac{\partial \Delta}{\partial \tau_{2}}+\left( \delta\omega_{1}-\delta\omega_{2} \right) \cos{\Delta} \cos{2 \theta} - \frac{1}{32} X \cos{2 \theta} ( \cos{\Delta} (3 \sigma_{11} \left(cos{\Delta} \sin{2 \theta}+1 \right)  \\ +3 \sigma_{22} \left( \cos{ \Delta} \sin{2 \theta}-1 \right) )  -2 \sigma_{12} \left( \cos{2 \Delta}+5 \right) \sin{2 \theta} ) = 0
\end{split}
\end{equation}
\end{widetext}

First of all, it is easy to show that equations \eqref{eq:NLSEangle} have two stationary points with coordinates ($\theta=\pi/4, \Delta = 0$) and  ($\theta = \pi/4, \Delta = \pi$). Taking into account the relations (\ref{Angle1}, \ref{psi}), one can observe that these points correspond to the steady states $\chi_{01}$ and $\chi_{02}$, respectively. 
All trajectories surrounding the stationary points describe the evolution of ''mixed'' states with different contributions of $\chi_{01}$ and $\chi_{02}$. 
In particular, the lines $\theta=0$ and $\theta=\pi/2$ correspond to the sum and the difference of the modes, respectively. 
Using the analogy between normal modes and coupled oscillators one can say that the stationary points mentioned take the role of the oscillators while the states with $\theta=0$ and $\theta=\pi/2$ are the analogies of in-phase and out-of-phase modes, respectively.
Only the difference is that the energies of in-phase and out-of-phase modes are equal.
Considering such states, one can see that the additional immovable points correspond to $\Delta=(\pi/2 \pm m\pi )$, where $m=0,1,2 \dots$.
On the other side these points lie in the trajectories, which separate the the normal modes attraction domains. 
Such a trajectory is the most distant from the stationary points and is named as the Limiting Phase Trajectory (LPT). 
As it was mentioned above the last notes the extremely nonuniform distribution of the energy (from a possible ones). 
 
\begin{figure}
a \includegraphics[width=35mm]{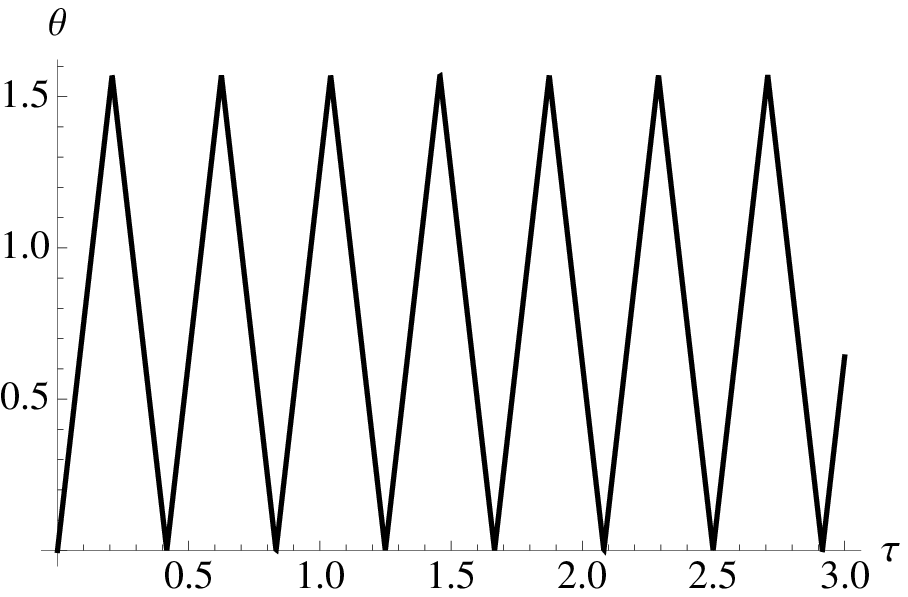} b \includegraphics[width=35mm]{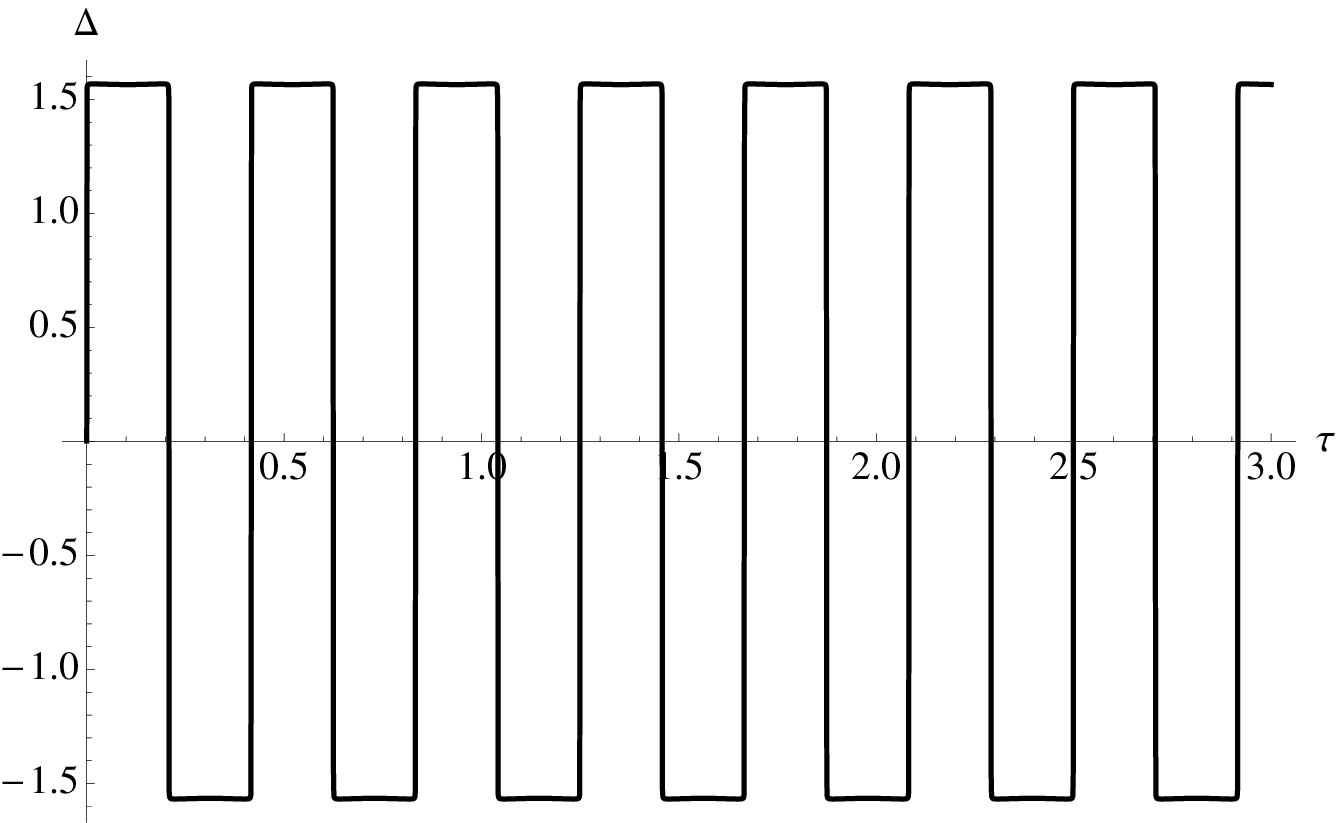} \\
c \includegraphics[width=35mm]{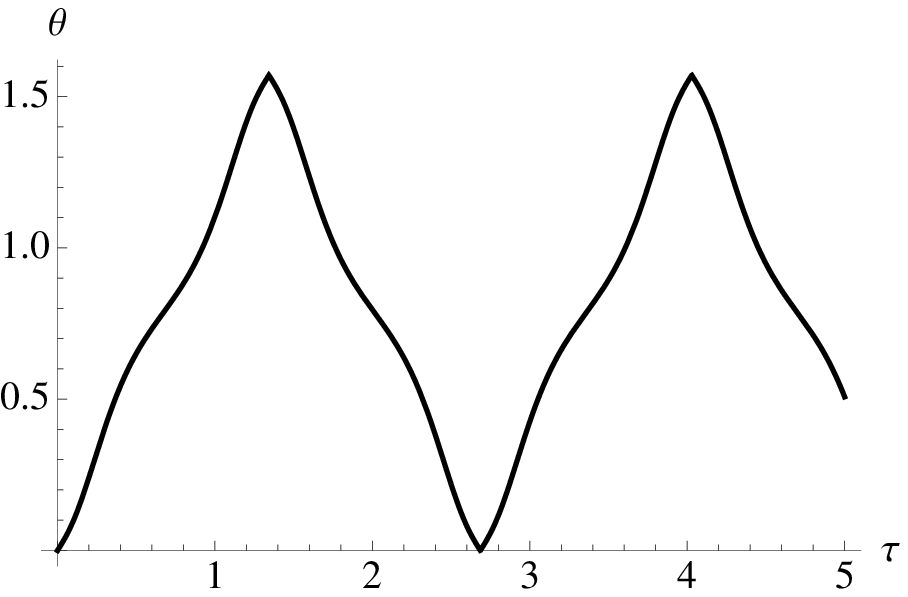} d \includegraphics[width=35mm]{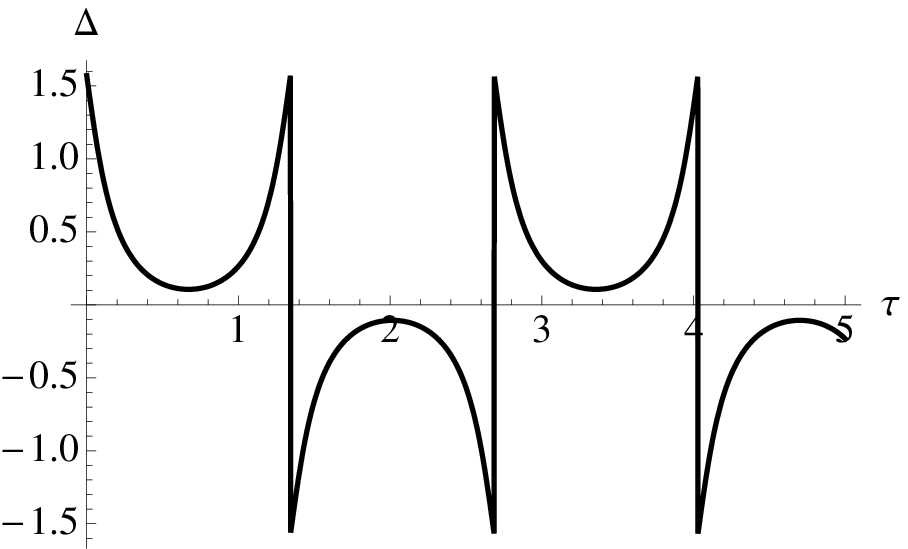} \\
e \includegraphics[width=35mm]{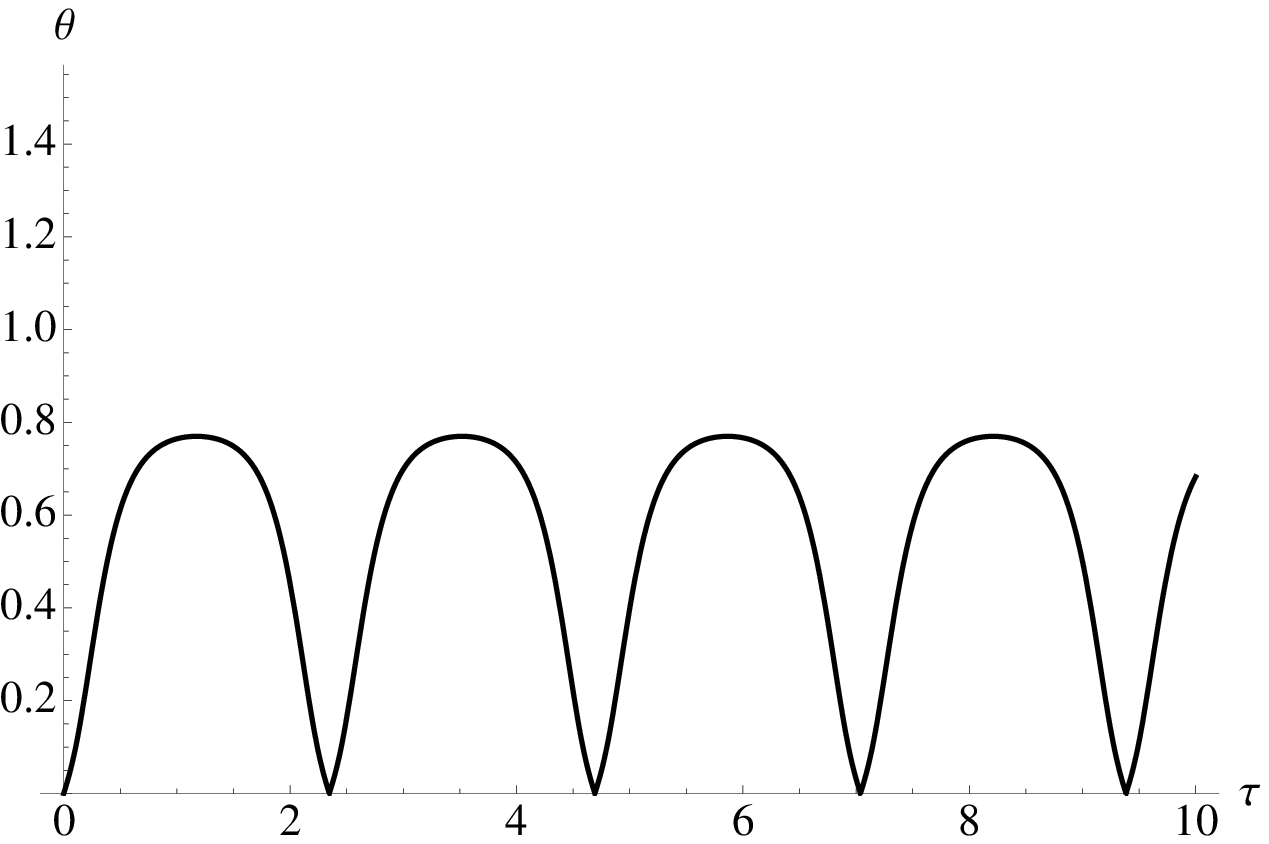} f \includegraphics[width=35mm]{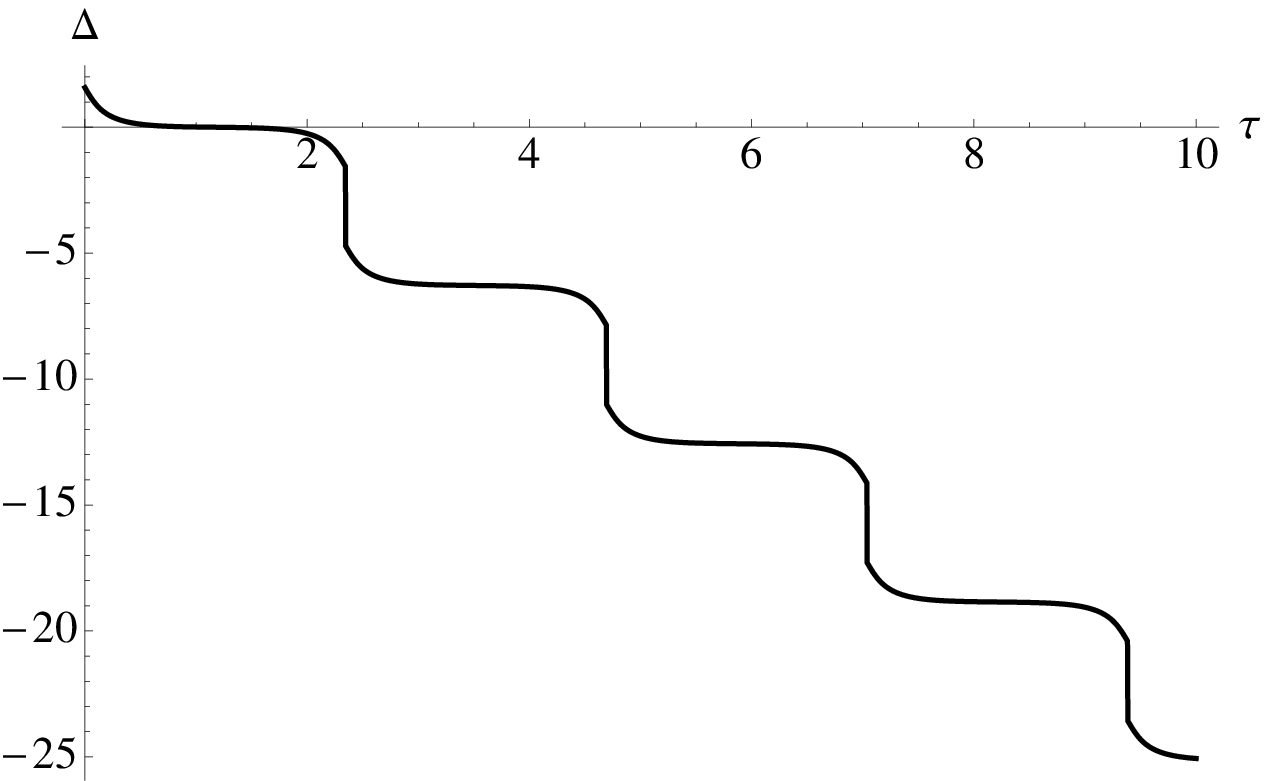}
\caption{Time evolution of the variables $\theta, \Delta$ for the different levels of the CNT excitation: (a-b) $X \ll X_{loc}$; (c-d) $X=0.995X_{loc}$; (e-f) $X=1.00005X_{loc}$}
\label{fig:TD_curves0}
\end{figure}

The numerical solutions of Eq. \eqref{eq:NLSEangle} with the initial conditions corresponding to the immovable point ($\theta=0$, $\Delta=\pi/2$) for the various values of the excitation $X$ are shown in the Fig. \ref{fig:TD_curves0} (a-f).

Figs. \ref{fig:TD_curves0}(a, b) show the evolution of $\theta(\tau_{2})$ and $\Delta(\tau_{2})$ for  small value of $X$, when the system is close to the linear one. 
In this case one can see the non-smooth behavior of the relative amplitudes as well as of the phase shift of the ''effective particles'' $\phi_{1}$ and $\phi_{2}$.
Such a behavior correlates with that the any states belonging to the lines $\theta=0$ or $\theta=\pi/2$ with $\Delta \ne( \pi/2 \pm m\pi )$, in fact, are some ''virtual'' ones, and they must be passed in the infinitesimal time.

Figs. \ref{fig:TD_curves0}(c,d) demonstrate the behavior of the functions $\theta$ and $\Delta$ if the excitation level $X$ is large enough and is extremely close, but smaller than some threshold value $X_{loc}$, the origin of which we will consider below. 
These figures show that qualitetively this behavior qualitatively does not different from that in fig. \ref{fig:TD_curves0}(a,b).

However, Figs.\ref{fig:TD_curves0}(e,f) exhibit the drastic changes in the evolution of the functions $\theta$ and $\Delta$ in spite of  excitation of the system was increased on $0.5 \%$ only. 
First of all the inteval of variation of the function $\theta$ becomes twice less. It is the most important fact, which shows, that the state with $\theta=0$ is inaccessible, if the initial conditions correspond to $\theta=\pi/2$ and vice versa. 
The second feature in the fig. \ref{fig:TD_curves0}(f) is the unlimited growth of the function $\Delta$. 
Such a behavior corresponds to the transit-time trajectories.

To clarify the variations of the solution of equations \eqref{eq:NLSEangle} let us rewrite the Hamilton function \eqref{eq:NLSEhamiltonian2} in terms of angular variables $\theta$ and $\Delta$ and consider the topology of the phase space.

\begin{multline}\label{eq:NLSEhamilton_angle}
H= \frac{X}{2} (  \delta \omega _{1} (\cos {\Delta } \sin {2 \theta }+1)  \\  +\delta \omega _{2} (1-\cos{\Delta } \sin{2 \theta } ) \\ -\frac{X}{64} ( \sigma _{11} (\cos{\Delta} \sin{2 \theta}+1)^{2}+  \sigma _{22} (1-\cos{\Delta} \sin{2 \theta } )^{2}  \\  +\sigma _{12} (\cos{2 \Delta } \cos{4 \theta } + 4 \sin^{2}{\Delta}+10 \cos^{2}{2 \theta } ) ) )
\end{multline}

\begin{figure*}
 \centering{
 (a) \includegraphics[ width=40mm]{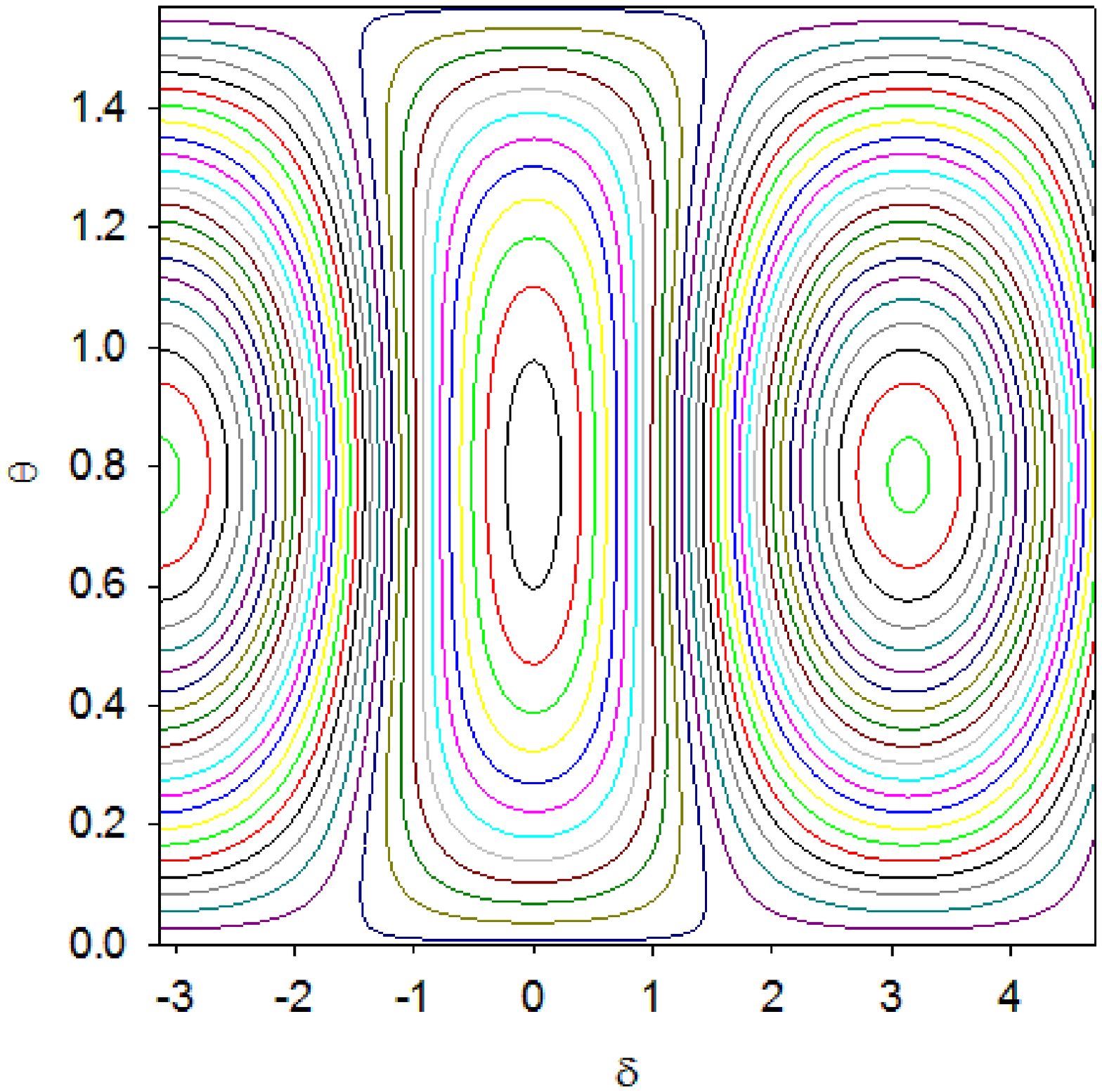}  \quad
 (b) \includegraphics[width=40mm]{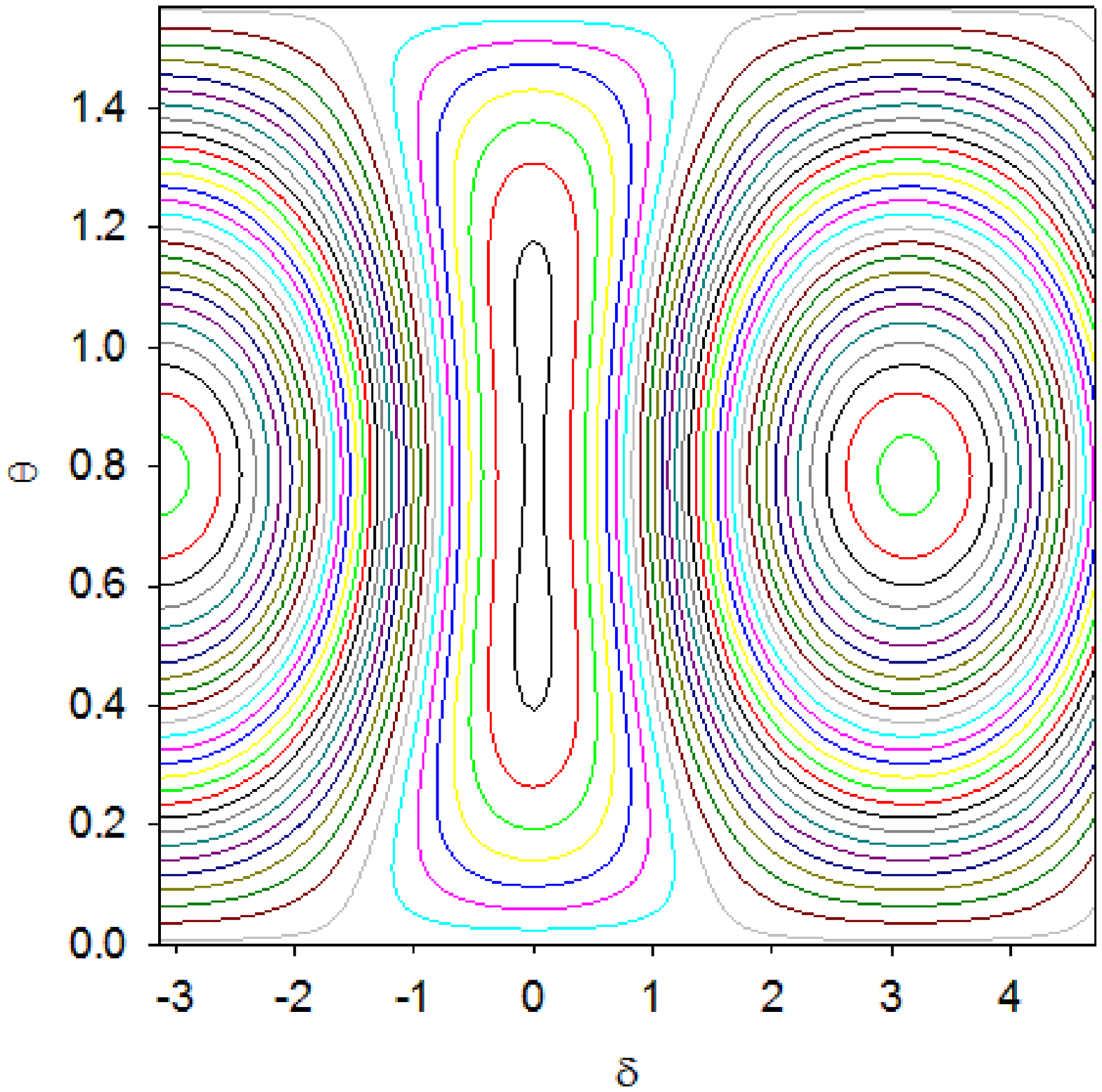} \quad 
(c) \includegraphics[width=40mm]{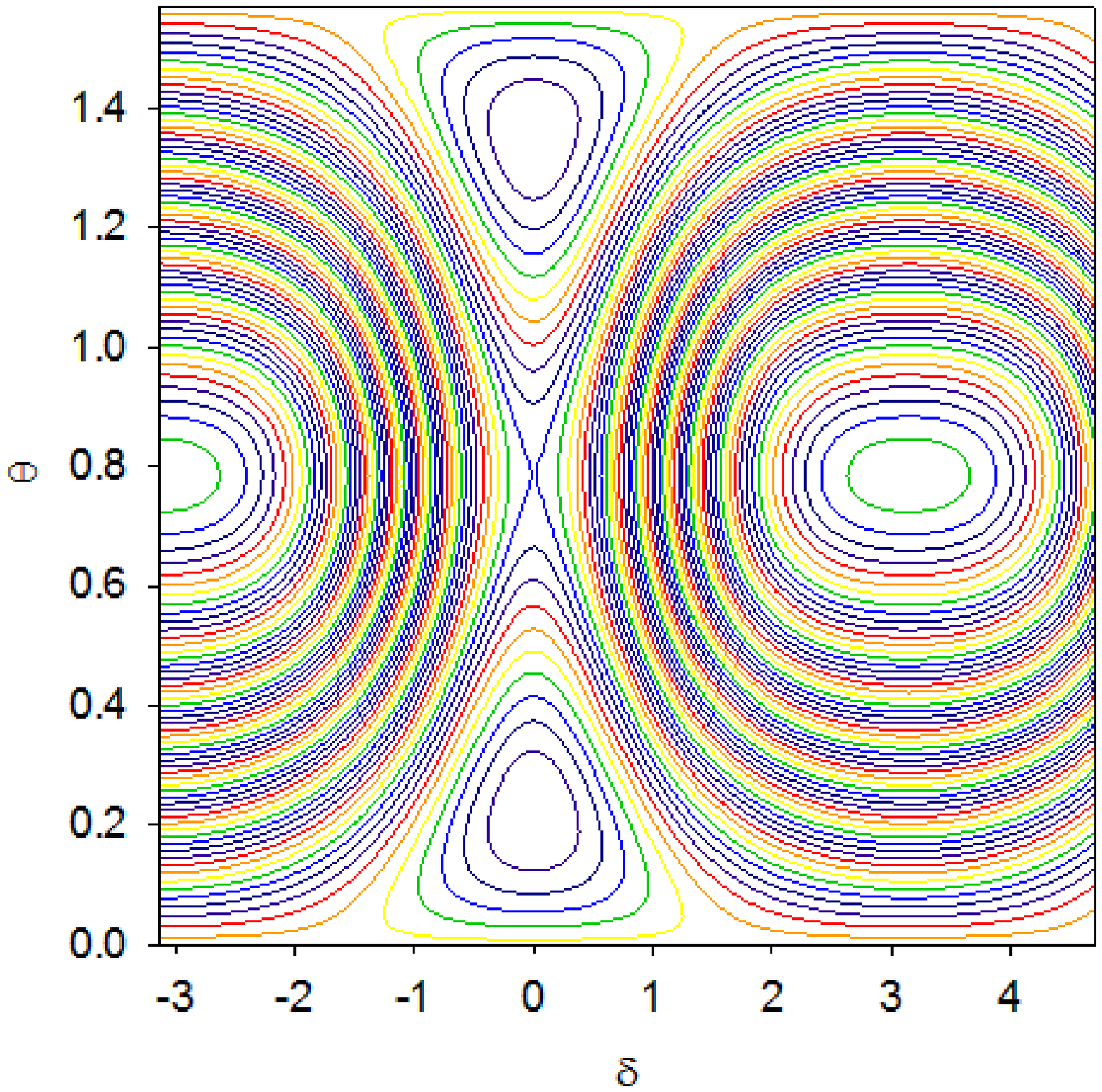}
}
\caption{(Color online) Phase portraits of the system with Hamiltonian \eqref{eq:NLSEhamilton_angle} for different values of the excitation: (a) $X \ll X_{loc}$, (b) $X=0.995 X_{loc}$, (c) $X=1.01X_{loc}$.}
\label{fig:NLSE_phaseportrait}
\centering{
(a) \includegraphics[width=40mm]{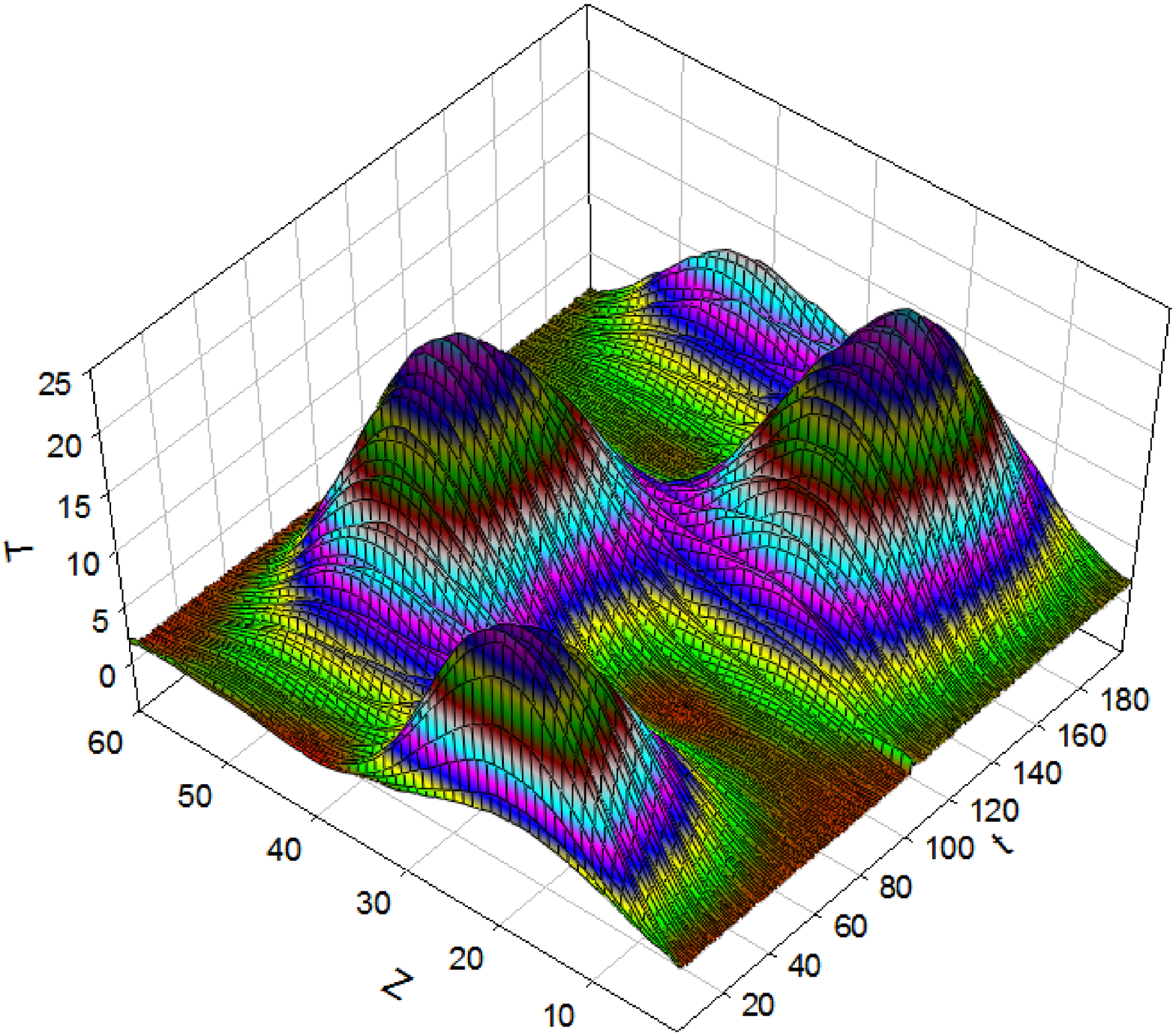} \quad
(b) \includegraphics[width=40mm]{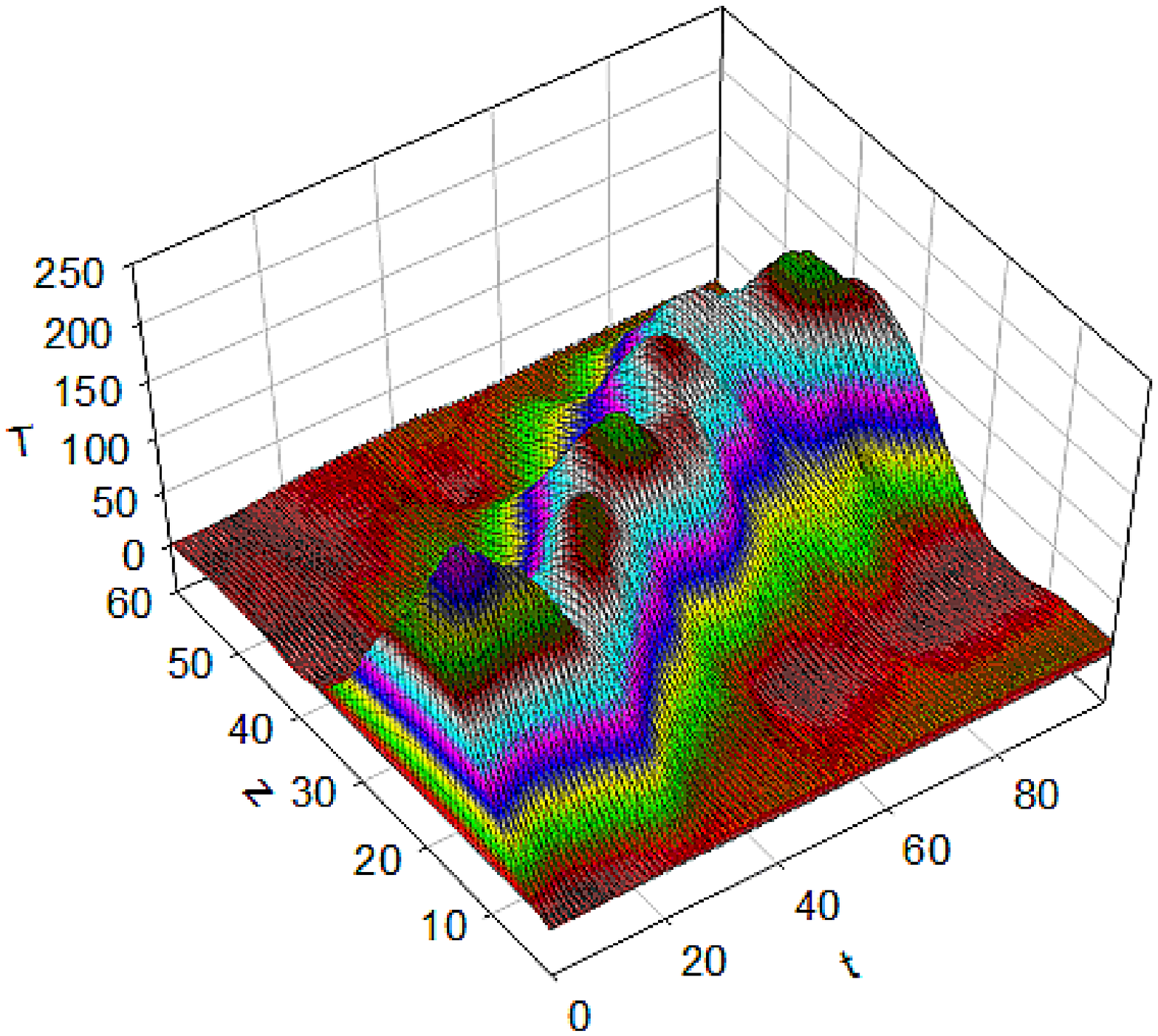}  \quad
(c)  \includegraphics[width=40mm]{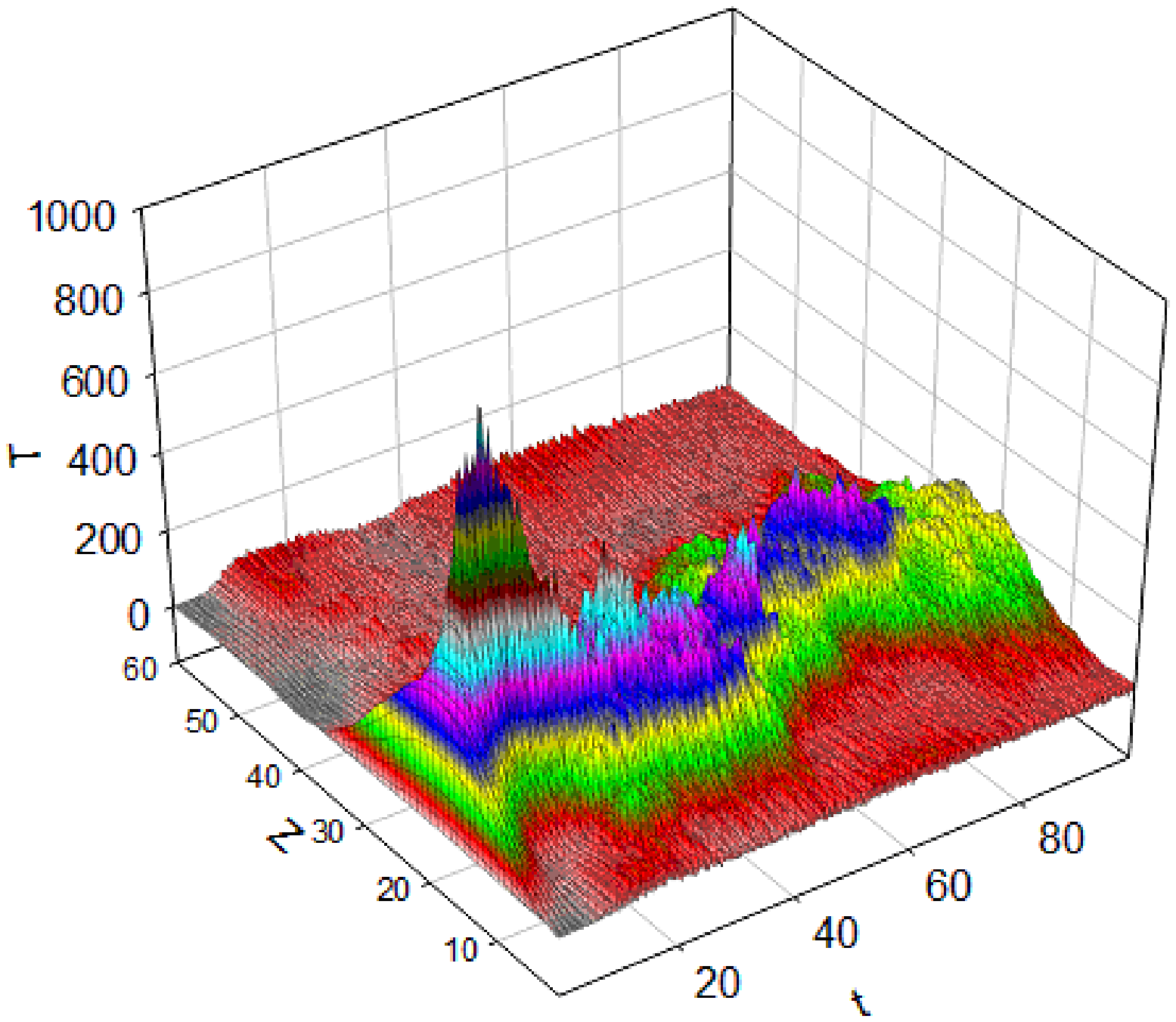}
}
\caption{(Color online) (a) The energy distribution along the CNT during the MD-simulation with the small excitation level $X$, that corresponded to phase portrait in the Fig. \ref{fig:NLSE_phaseportrait}(a). (b - c) The same as in the panel (a) with initial excitation level $X \lesssim X_{loc}$  and $X > X_{loc}$, respectively. }
\label{fig:MDenergy}
\end{figure*}

Fig. \ref{fig:NLSE_phaseportrait} shows the phase portraits for various values of the parameter $X$.
 The initial structure of the phase space for the small $X$, when the system is close to the linear, is clearly seen in the left panel of Fig. \ref{fig:NLSE_phaseportrait}. 
 The representative domains of the phase space are bounded by the intervals $0 \le \theta \le \pi/2$ and $-\pi/2 \le \Delta \le 3 \pi/2$.
  Two stable stationary points correspond to the normal modes $\chi_{01}$ and $\chi_{02}$.
  The trajectory, which separates the attraction domains of different steady states and rounds the normal mode $\chi_{01}$, contains two lines $\theta = 0$ and $\theta=\pi/2$, and two fragments, which connect the pairs of points: (($\theta=0, \Delta=-\pi/2$); ($\theta=\pi/2, \Delta=-\pi/2$)) and (($\theta=0, \Delta= \pi/2$); ($\theta=\pi/2, \Delta=\pi/2$)).
  The analogous trajectory rounds the steady state $\chi_{02}$.
  The motion along these trajectories leads to the non-smooth behavior as it is shown in the Fig. \ref{fig:TD_curves0}
  
 However, the steady state $\chi_{01}$ becomes unstable if the parameter $X$ exceeds some threshold. Its value $X_{ins}$ can be calculated from the condition of the instability:
 
 \begin{equation}\label{eq:NLSE_inst}
\begin{split}
 \frac{\partial^{2} H}{\partial \theta^{2}}_{| \{\Delta=0, \theta=\pi/4 \} } = 0,  \\
 X_{ins}=\frac{16 \left(\delta \omega _{1}-\delta \omega _{2}\right)}{\sigma _{11}-6 \sigma _{12}}.
\end{split}
\end{equation}
 
 Two new stationary points arise after loosing the mode $\chi_{01}$ its stability.
 They are new nonlinear normal modes.
 The distance between them grows while the parameter $X$ increases. 
 These new stationary points correspond to some non-uniform distribution of the energy along the CNT, however, this non-uniformity is weak. 
 The main features of these states consists in that no trajectory surrounding them cannot attain the separatrix, which passes through the unstable stationary state $\chi_{01}$. 
 Therefore, the non-uniformity of energy distribution remains for the infinite time. 
 Nevertheless, any trajectories, which are situated in the gap between the separatrix and the LPT, preserve the possibility to pass from the vicinity of $\phi_1$ state ($\theta = 0$) into the vicinity of $\phi_{2}$ state ($\theta = \pi/2$) (see Fig. \ref{fig:TD_curves0}(c, d). 
 These process is accompanied with the slow energy transfer from one part of the CNT to another one.
 
 However, the behavior of the solution of equations \eqref{eq:NLSEangle} is changed drastically if the value of $X$ overcomes next threshold $X_{loc}$. 
 The existence of this threshold results from that the new stationary states move away from the unstable state and the separatrix grows while the LPT moves to the unstable state in the vicinity of $\theta=\pi/4$.
 The principal changes happen when the LPT reaches the point ($\theta=\pi/4, \Delta = 0$).
 At this moment the gap between the LPT and the separatrix disappears and the only trajectory passed from $\theta=0$ to $\theta=\pi/2$ is the LPT. 
 The further increasing of the parameter $X$ leads to that new separatrix, which is passed through the unstable stationary points $(\theta=\pi/4, \Delta=0)$ and $(\theta=\pi/4, \Delta=2 \pi)$, arises (see Fig. \ref{fig:NLSE_phaseportrait}(c)).
 It separates the phase space of the system into uncoupled parts and any trajectories, which start near the $\theta = 0$, cannot attain the value $\theta= \pi/2$ and vice versa. 
 It means that the energy originally given to in a part of CNT is kept in this part. 
 The new LPTs enclose the stationary points, which correspond to the stable nonlinear modes.
 Fig. \ref{fig:NLSE_phaseportrait}(c) shows the transit-time trajectories, which are in the domain between LPTs and separatrix.
 Therefore the solution, which is shown in the Fig. \ref{fig:TD_curves0}(f), demonstrates the infinite rise of the variable $\Delta$.
 
 The condition of the bifurcation discussed is the degeneration of the energy of the states $\chi_{01}$, $\phi_{1}$ and $\phi_{2}$, i.e.
 
 \begin{equation}\label{eq:NLSElocalization}
 \begin{split}
 H(\theta=0, \delta=0) =  H(\theta=\pi/4, \delta=0)  \\
 =  H(\theta=\pi/2, \delta=0).
 \end{split}
 \end{equation}
 
 So, the value of the localization threshold turns out to be
 
 \begin{equation}\label{eq:NLSEXloc}
 X_{loc}=\frac{64 \left(\delta \omega _1-\delta \omega _2\right)}{3 \sigma _{11}-\sigma _{22}-12 \sigma _{12}}
 \end{equation}
 
 \begin{figure}
 \includegraphics[width=60mm]{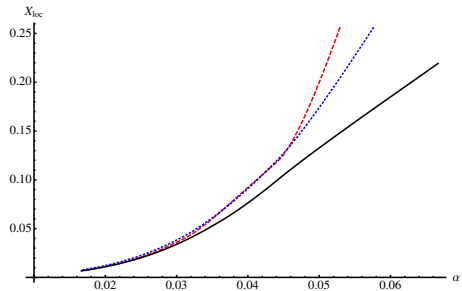}
 \caption{The threshold of vibration localization for the CNT with radius 0.79 nm vs inverse aspect ratio $\alpha$: solid black and the dashed red curves are the pedictions, which based on the Eqs. \eqref{eq:NLSE_mod}  and \eqref{eq:NLSE}, respectively. The dotted blue curve shows the threshold value estimated by the numerical method (see section \ref{sec:Num} for detail). }
 \label{fig:Xloc1}
 \end{figure}
   
   Fig. \ref{fig:Xloc1} demonstrates the dependence of the localization threshold  in the terms of the radial displacement $w$ from the inverse aspect ratio of the CNT. 
 The solid curve shows the threshold value in the accordance with equation \eqref{eq:NLSEXloc}. 
 To compare with it, the threshold, which has been estimated on the base of equation \eqref{eq:NLSE}, is drawn by the dashed line.
  One can see that the asymptotic values for the long CNTs (at $\alpha \rightarrow 0$) of both thresholds are the same, but the difference between them becomes essential for the finite-length CNT.

\subsection{Comparison of two-mode approximation with other numerical methods}\label{sec:Num}

The two-mode approximation in the framework of nonlinear Sanders-Koiter theory of thin elastic shells allows us to predict the bifurcation of dynamical behavior of low-frequency CNT vibrations as well as to estimate the threshold values of oscillation amplitude. 
However, the influence of the other part of the spectrum is very important for the estimation of the reliability of the obtained results. 
Therefore, they should be verified by the independent numerical methods.
One of the approaches consists in the direct numerical integration of the modal nonlinear equations of the Sanders-Koiter thin shell theory. 

In order to carry out the numerical analysis of the CNT dynamics, a two-step procedure was used: i) the displacement field was expanded by using a double mixed series, then the Rayleigh-Ritz method was applied to the linearized formulation of the problem, in order to obtain an approximation of the eigenfunctions; ii) the displacement fields are re-expanded by using the linear approximated eigenfunctions, the Lagrange equations were then considered in conjunction with the nonlinear elastic strain energy to obtain a set of nonlinear ordinary differential equations of motion.

So, to satisfy the boundary conditions the displacement field was expanded into series 

\begin{equation}\label{eq:SKT2x}
\begin{split}
r(\xi ,\varphi,t )=\left[ \sum _{m=0}^{M_{u} }\sum _{n=0}^{N}R_{m,n} T_{m}^{*} (\xi )\cos{n\varphi} \right] f(t) \\
\end{split}
\end{equation}

where function $r(\xi, \varphi,t)$ substitutes the displacements $u$, $v$ or $w$. 

In equations \eqref{eq:SKT2x}  $T_{m}^{*} (\xi )=T_{m} (2\xi -1)$ are the Chebyshev orthogonal polynomials of the $m-th$ order, $n$ is the number of nodal diameters, and $f(t)$ describes the time evolution of the CNT vibrations.

The maximum number of variables needed for describing a general vibration mode with $n = 2 $ nodal diameters (Circumferential Flexural Mode) is obtained by the relation ($N_{max} = M_{u} + M_{v} + M_{w} + 3 - p $ ), where ( $M_{u} = M_{v} = M_{w} $) denote the maximum degree of the Chebyshev polynomials and $ p $ describes the number of equations for the boundary conditions to be respected.

A specific convergence analysis was carried out to select the degree of the Chebyshev polynomials: degree 11 was found suitably accurate, ( $ M_{u} = M_{v} = M_{w} = 11 $).

In the cases of a SWCNT with simply supported or clamped edges ( $p  = 8 $), the maximum number of degrees of freedom of the system with is equal to ( $N_{max} = 33 + 3 - 8 = 28 $).

Conversely, in the case of a SWNT with free edges ( $p = 0 $), the maximum number of degrees of freedom of the system is equal to ( $N_{max} = 33 + 3 = 36 $).

The equations \eqref{eq:SKT2x} are inserted into the expressions of the potential energy $E_{el}$and kinetic energy $T$ to compute the Rayleigh quotient $ R(q)=E_{el, \max} / T^{*} $, where $E_{el, max } =\max (E_{el})$ is the maximum of the potential energy during a modal vibration, $T^{*} =T_{\max } /\omega ^{2} $, $T_{\max } =\max (T)$ is the maximum of the kinetic energy during a modal vibration, $\omega$ is the circular frequency of the synchronous harmonic motion and $ q =[...,U_{m,2} ,V_{m,2} , W_{m,2} ,...]^{T} $ represents a vector containing all the unknown variables.

After imposing the stationarity to the Rayleigh quotient, one obtains the eigenvalue problem

\begin{equation}\label{eq:SKT8}
(-\omega ^{2} M+ K)  q = 0
\end{equation}

which gives approximate natural frequencies (eigenvalues) and modes of vibrations (eigenvectors).
The results of performed calculation show that the eigenspectrum values are in the good accordance with the estimations made in the framework of reduced Sanders-Koiter theory discussed above.
 The specific difference between eigenvalues amounts to the values $2 - 4\%$ for the long-wave modes and reachs up to $20\%$ while the longitudinal wavenumber grows \citep{SoundVibr2014}.

In the nonlinear analysis, the full expression of the dimensionless potential energy $E_{el}$ containing terms up to the fourth order (cubic nonlinearity), is considered.

In the cases of simply supported and clamped boundary conditions, the two low-frequency optical-type circumferential flexure modes ($m=1, n=2$) and ($m=2, n=2$) are considered.

Using the Lagrange equations

\begin{equation}\label{eq:SKT14}
\frac{d}{d\tau } \left(\frac{\partial T}{\partial q^{'} _{i} } \right)+\frac{\partial E_{el}}{\partial q_{i} } =0,
\end{equation}

a set of nonlinear ordinary differential equations is obtained; these equations must be completed with suitable initial conditions on displacements and velocities. This system of nonlinear equations of motion was finally solved by using the implicit Runge-Kutta numerical methods with suitable accuracy, precision and number of steps. The solution of nonlinear equations with initial conditions in the vicinity of the bifurcation threshold shows the coincidence of the threshold values in the analytical model and the numerical one for the wide interval of aspect ratios (see Fig. \ref{fig:Xloc1}). The procedure and results will be discussed in the nearest future.

\subsection{MD simulation}

\begin{figure}
\includegraphics[width=70mm]{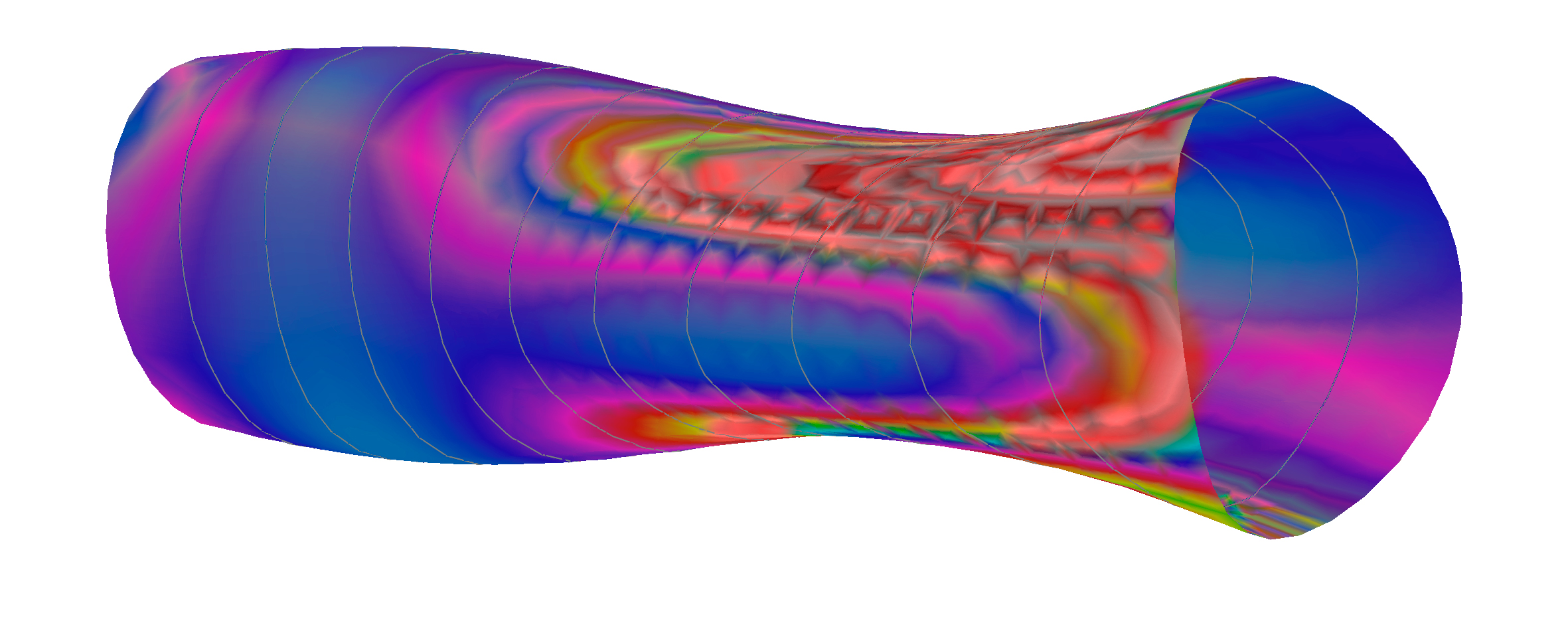}
\caption{(Color online) Snapshot of typical energy distribution along the deformed CNT during the vibration associated with CFM spectrum branch.}
\label{fig:CNT_deform}
\end{figure}

To verify the results of analytical model the simulation of the low-frequency vibrations of CNTs was performed by  molecular dynamics (MD)  techniques using realistic inter-atomic  potential  functions. 
Classical molecular dynamics technique which uses predefined potential functions (force fields) was applied for the calculation of  the total potential energy of  the system.  
The typical MD experiment consisted of several stages. 
At the first stage the CNT was kept at  high temperature ($\simeq 400 K$) for structural relaxation. 
Then the termostat temperature was decreased with a constant rate down to approximately 1 K with a subsequent low-temperature relaxation. 
The third stage dealt with CNT deformation according to analytical solution with subsequent relaxation. 
The second version of initial conditions were given by initial velocities of atoms at zero initial displacements. After that the external field was turned off, and the free natural oscillations of CNT with the fixed boundary conditions were realized. In accordance with analytical description, the atoms at the edges of CNT were fixed by the force field against any radial displacements ($W(0,t)=W(1,t)=0$) that corresponds to the boundary conditions similar to simply supported shell. The typical snapshot of distribution of the CNT deformation energy during the MD simulation is shown in fig.(\ref{fig:CNT_deform}). 

The consequent analysis of MD simulation data included  the control of  natural frequencies and energy distribution along the CNT axis via  variation of the oscillation amplitude. The 3D pictures of energy distribution along the CNT axis measured during the MD simulations have been discussed in the section \ref{s:LPT} (see the figures \ref{fig:MDenergy} (a-c)).

\begin{figure}
\includegraphics[width=70mm]{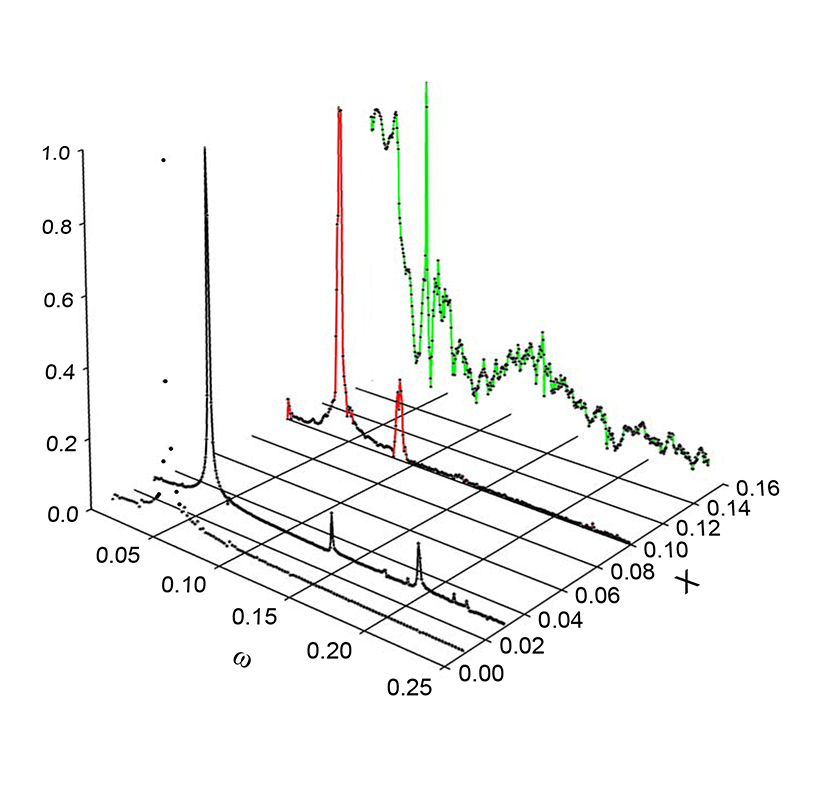} \quad
\caption{(Color online) CNT vibration spectra at the various excitation level: $X=(0.01 - 1.1)X_{loc}$.}
\label{fig:MDspectra}
\end{figure}

Fig. \ref{fig:MDspectra} shows the variation of the CNT vibration spectrum with changing of the initial excitation level. 
The dot, black and red curves correspond to the excitation $X < X_{loc}$. 

One should note, that the large narrow peak near the frequency $\omega \simeq 0.05$ corresponds to the lowest eigenvalue of the system under consideration. 
Therefore no vibration with the smaller frequency exists in the gap for the any initial excitation, if they do not exceed the localization threshold $X_{loc}$. 
However, the overcoming the threshold of localization changes the spectrum drastically. 
From one side, one can see in Fig. \ref{fig:MDenergy}(c), that the shape of the initial excitation is deformed essentially during the MD simulation and the local temperature reaches a great value ($\sim 1000 K$). 
Therefore, the onset of high-frequency modes is naturally sufficient.
On the other side, the intensive oscillations in the gap of the spectrum can not be explained by the increasing of the local temperature. To fill this gap, the excitation of another low-frequency modes is required (they may be acoustic type modes like the bending or the longitudinal stretching modes - see Fig. \ref{fig:CNT2chain}). 
The another possibility is forming of the localized excitation, because the Fourier spectrum is wide enough.
Unfortunately, the two-mode approximation used in our analysis can not answer in this question. 
The accurate study of this problem needs in the consideration of the inter-brunch mode interaction.
This problem will be formulated in our future studies.

\section{Conclusion}\label{conclusion}


 Instability and bifurcation of the edge-spectrum optical NNM at the value $X = X_{ins}$  leads to appearance of the localized NNMs with stationary (in slow time scale) energy localization in some part of the CNT. 
 In contrast to a breather,  this is not  a strong localization because the two-mode approximation, which is valid for considered aspect ratio, can reveal  a weak  localization only. 
 
We demonstrate that instability of edge-spectrum optical modes of CNT vibrations is the preliminary condition of \textit{non-stationary (in slow time scale)} energy localization in the some domain of CNT. 
The energy capture in one of the CNT parts can be achieved, if the excitation level exceeds the specific threshold $X=X_{loc}$, which corresponds to merging two trajectories, which are the LPT and the separatrix appeared at $X=X_{ins}$. 
When this  threshold is exceeded  the phase portrait of the system under consideration changes drastically: the separatrix passing through the unstable stationary point ($\theta=\pi/4$, $\Delta=0$) (see fig. \ref{fig:NLSE_phaseportrait}(b)) encircles the stable stationary point ($\theta=\pi/4$, $\Delta=\pi$) and prevents full energy exchange between effective particles $\psi_{1}$ and $\psi_{2}$. 
Simultaneously a set of transit-time trajectories, which involve any values of phase difference $\Delta$, is created. 
It means that initial conditions corresponding to identical velocities or displacements  of both  modes lead to the energy capture by the effective particles. 
Then only a partial energy exchange becomes possible along the trajectories, surrounding the stable stationary point and situated inside the separatrix. 
One should keep in mind that the process of energy capture does not suggest the creation of strongly localized solutions whose formation requires a participation of more  components of the spectrum. 
This can be achieved for CNT with larger aspect ratio.

It should be noted once more that the development and the use of the analytical framework based on the LPT concept is motivated by the fact that resonant non-stationary processes occurring in a broad variety of finite dimensional physical models are beyond the well-known paradigm of nonlinear normal modes (NNMs), fully justified only for quasi-stationary processes and non-stationary processes in non-resonant case. While the NNMs approach has been proved to be an effective tool for the analysis of instability and bifurcations of stationary processes (see, e.g., \citep{Vak96}), the use of the LPTs concept provides the adequate procedures for studying strongly modulated regimes as well as the transitions to energy localization and chaotic behavior \citep{VVS2010}. Such an approach clarifies also the physical nature of the breathers formation in infinite discrete or continuum systems.

As a conclusion we would like to note that the phenomenon of energy localization considered above has universal character and it is the common peculiarity of the systems possessing the optical-type branches of vibrational spectrum. 
However, as it was studied early \citep{VVS2010, VVS2011, DAN2010}, the occurence of the localization depends on the types of nonlinearity as well as on the relations between coefficients $\sigma_{i,j}$ in the Hamiltonian \eqref{eq:NLSEhamiltonian2}. 
If some ratios between coefficients $\sigma_{i,j}$ are sa tisfied, an additional integral of motion arises that leads to the effective linearization of the equations of motion and, as consequence, to the absence of the localization processes \citep{CISM2010}. 
So, in spite of that the interaction of resonating nonlinear modes describes by the Hamiltonian \eqref{eq:NLSEhamiltonian2} for a wide class of the nonlinear systems, the results of this interaction may vary considerably. 
In any case, the analysis of the Hamiltonian \eqref{eq:NLSEhamiltonian2} in combination with the LPT concept gives us usefull tool for the study of nonlinear systems.


\begin{acknowledgments}
The work was supported by Russia Basic Research Foundation (grant 08-03-00420a) and Russia Science Foundation (grant 14-17-00255)
\end{acknowledgments}

\appendix

\section[level(1)]{The reduced Sanders- Koiter thin shell theory}\label{app:A}

It is convenient to use the dimensionless variables which determine the elastic deformation of circular thin shell. 
In such a case all components of the displacement field ($u$ - longitudinal along the CNT axis, $v$ - tangential and $w$ - radial displacement, respectively) are measured in the units of CNT radius $R$. 
The displacements and respective deformations refer to the middle surface of the shell. 
The coordinate along the CNT axis  $\xi=x/L$ is measured via the length of nanotube and variates from $0$ up to $1$, and $\theta$ is the azimuthal angle. 

One can define the dimensionless energy and time  variables, which are measured in the units $E_{0}=YRLh/(1-\nu^{2})$ and $t_{0}=1/\sqrt{Y/\rho R^{2}(1-\nu^{2})}$, respectively. 
Here $Y$ is the Young modulus of graphene sheet, $\rho$ - its mass density, $\nu$ - the Poisson ratio of CNT, and $h$ is the effective thickness of CNT wall. There are two dimensionless geometric parameters which characterize CNT: the first one is inverse aspect ratio $\alpha=R/L$ and the second - effective thickness shell $\beta=h/R$.

The energy of elastic deformation of CNT in the dimensionless units is written as follows:

\begin{equation}\label{energy_elst}
\begin{split}
E_{el}=\frac{1}{2}\int\limits_{0}^{1}\int\limits_{0}^{2 \pi} ( N_{\xi}\varepsilon_{\xi}+N_{\varphi}\varepsilon_{\theta}+N_{\xi \varphi}\varepsilon_{\xi \varphi} + \\ M_{\xi}\kappa_{\xi}+M_{\varphi}\kappa_{\varphi}+M_{\xi \varphi}\kappa_{\xi \varphi} ) d\xi d\varphi
\end{split}
\end{equation}

where $\varepsilon_{\xi}$, $\varepsilon_{\varphi}$ and $\varepsilon_{\xi \varphi}$ are the longitudinal, circumferential and shear deformations, and $\kappa_{\xi}$, $\kappa_{\varphi}$ and $\kappa_{\xi \varphi}$ are the longitudinal and circumferential curvatures, and torsion, respectively. 
The respective forces and momenta may be written in the physically linear approximation:

\begin{equation}\label{eq:forces}
\begin{split}
N_{\xi}=\varepsilon_{\xi}+\nu \varepsilon_{\varphi}, \quad N_{\varphi}=\varepsilon_{\varphi}+\nu \varepsilon_{\xi}, \\
\quad N_{\xi \varphi}=\frac{1-\nu}{2} \varepsilon_{\xi \varphi} \\
M_{\xi}=\frac{\beta}{12}(\kappa_{\xi}+\nu \kappa_{\varphi}), \quad M_{\varphi}=\frac{\beta}{12}(\kappa_{\varphi}+\nu \kappa_{\xi}), \\
\quad M_{\xi \varphi}=\frac{\beta}{24}(1-\nu) \kappa_{\xi \varphi}
\end{split}
\end{equation}

 One should note that both curvatures and torsion are the dimensionless variables in accordance with our definition of dispacement field $(u, v, w)$.

The Sanders-Koiter approximation of defectless thin shell  allows to write the nonlinear deformations $(\varepsilon)$  and curvatures  $(\kappa)$  in the following form

\begin{equation}\label{deformation}
\begin{split}
\varepsilon_{\xi} = \alpha \frac{\partial u}{\partial \xi} + \frac{\alpha^{2}}{2}( \frac{\partial w}{\partial \xi})^{2} +\frac{1}{8}(\alpha \frac{\partial v}{\partial \xi}-\frac{\partial u}{\partial \varphi})^{2} \\ 
\varepsilon_{\varphi} = \frac{\partial v}{\partial \varphi} + w + \frac{1}{2} (\frac{\partial w}{\partial \varphi} -v)^{2}+\frac{1}{8}(\frac{\partial u}{\partial \varphi}-\alpha \frac{\partial v}{\partial \xi})^{2}  \\
 \varepsilon_{\xi \varphi} = \frac{\partial u}{\partial \varphi} + \alpha \frac{\partial v}{\partial \xi} +\alpha \frac{\partial w}{\partial \xi}(\frac{\partial w}{\partial \varphi}-v) 
\end{split}
\end{equation}

\begin{equation}\label{curvation}
\begin{split}
\kappa_{\xi} = - \alpha^{2} \beta \frac{\partial^{2} w}{\partial \xi^{2}}, \\ 
\kappa_{\varphi} = \beta \left(\frac{\partial v}{\partial \varphi} - \frac{\partial^{2} w}{\partial \varphi^{2}} \right), \\ 
\kappa_{\xi \varphi} = \beta \left( - 2 \alpha \frac{\partial^{2} w}{\partial \xi \partial \varphi} + \frac{3 \alpha}{2} \frac{\partial v}{\partial \xi} - \frac{1}{2} \frac{\partial u}{\partial \varphi}\right).
\end{split}
\end{equation}
  
    One should make some physically grounded relationships between the displacement components  to simplify the description of the CNT nonlinear dynamics. 
    We consider the low-frequency optical-type vibrations which are specified by circumferential wave number $n=2$. This branch is characterized by relatively small circumferential and shear deformations, while the displacements themselves may not be small. In such a case we can write:

 \begin{equation}\label{eq:hypot}
 \varepsilon_{\varphi}=0 ; \varepsilon_{\xi \varphi}=0
 \end{equation}
The components of displacement field are 

\begin{equation}\label{eq:var1}
\begin{split}
u(\xi, \varphi, \tau)=U_{0}(\xi,\tau)+U(\xi,\tau) \cos( n\varphi) \\
v(\xi, \varphi, \tau)=V(\xi,\tau) \sin(n\varphi) \\
w(\xi, \varphi, \tau)=W_{0}(\xi,\tau)+W(\xi,\tau) \cos(n \varphi)
\end{split}
\end{equation}

These relations allow us to express the longitudinal and tangential components, and axially symmetric part of the radial displacement via the radial one. Corresponding relationships can be written as folows:

\begin{equation}\label{eq:uv2w}
\begin{split}
V(\xi,\tau)=-\frac{1}{n} W(\xi,\tau);  \\ 
U(\xi,\tau)=-\frac{\alpha}{n^{2}} \frac{\partial W(\xi,\tau)}{\partial \xi} \\
W_{0}(\xi,\tau)=-\frac{1}{4 n^{2}}((n^{2}-1)^{2} W^{2}(\xi,\tau)+\alpha^{2} (\frac{\partial W(\xi,\tau)}{\partial \xi})^{2});  \\ 
\frac{\partial U_{0}(\xi,\tau)}{\partial \xi}=-\frac{n^{2}+1}{4 n^{2}} \alpha (\frac{\partial W(\xi,\tau)}{\partial \xi})^{2}
\end{split}
\end{equation}

Because the  kinetic energy contains the inertial terms corresponding to all components of the deformation field

\begin{equation}\label{energy_kin}
\begin{split}
E_{kin} =\frac{1}{2} \int\limits_{0}^{1} \int\limits_{0}^{2 \pi}\left((\frac{\partial u}{\partial \tau})^{2} + (\frac{\partial v}{\partial \tau})^{2} + (\frac{\partial w}{\partial \tau})^{2}\right)d\xi d\varphi  
\end{split}
\end{equation}

 we need in  taking into account the relations \eqref{eq:uv2w} also.

Omitting the calculation details one can write the final equation of motion in terms of radial displacement $W(\xi, t)$:

\begin{widetext}
\begin{multline}\label{eq:nonlin}
 \frac {\partial^{2}W}{\partial \tau^{2}}+  \omega_{0}^{2} W -\mu \frac{\partial^{2} W}{\partial \xi^{2}} -  \gamma \frac{\partial ^{4} W}{\partial \xi^{2} \partial \tau^{2}}+\kappa \frac{\partial^{4} W}{\partial \xi^{4}}   
 + a_{1} W \left( \left( \frac{\partial W}{\partial \tau} \right) ^{2}+W \frac{\partial ^{2} W}{\partial \tau^{2}} \right)     
 + a_{2} \left( \frac{\partial W}{\partial \xi} \right)^{2} \frac{\partial ^{2} W}{\partial \xi^{2}} +  \\ 
  a_{3} \left( 2 \frac{\partial W}{\partial \tau}\frac{\partial W}{\partial \xi} \frac{\partial ^2 W}{\partial \xi \partial \tau} -W \left( \frac{\partial^{2} W}{\partial \xi \partial \tau} \right)^{2}+ \left( \frac{\partial W}{\partial \tau} \right)^{2} \frac{\partial^{2} W}{\partial \xi^{2}} \right)  \\
 +a_{4} \left[ \frac{\partial ^{2} W}{\partial \xi \partial \tau} \left( \frac{\partial ^{2} W}{\partial \xi \partial \tau} \frac{\partial ^{2} W}{\partial \xi ^{2}}+2 \frac{\partial W}{\partial \xi } \frac{\partial ^{3} W}{\partial \xi^{2} \partial \tau} \right) + \frac{\partial W}{\partial \xi } \left( \frac{\partial W}{\partial \xi} \frac{\partial ^{4} W}{\partial \xi^{2} \partial \tau^{2}}+ 2 \frac{\partial ^{3} W}{\partial \xi \partial \tau^{2}}\frac{\partial ^{2} W}{\partial \xi^{2}} \right) \right]=0 
 \end{multline}
 \end{widetext}
 
where

\begin{equation}\label{eq:param}
\begin{split}
\omega_{0}^{2}=\beta^{2} \frac{n^{2} (n^{2}-1)^{2}}{12(n^{2}+1)}, \quad
\mu=\alpha^{2} \beta^{2} \frac{ (n^{2}-1)(n^{2}-1+\nu)}{6 (n^{2}+1)};  \\ 
\gamma = \frac{\alpha^{2}}{n^{2}(n^{2}+1)},  \quad \quad  \kappa =\frac{\alpha^{4} (12+n^{4} \beta^{2})}{12 n^{2}(n^{2}+1)} \thicksim  \frac{\alpha^{4} }{n^{2}(n^{2}+1)};   \\
a_{1} = \frac{(n^{2}-1)^{4}}{2 n^{2} (n^{2}+1)},  \quad \quad
a_{2} = 2 \alpha^{4}  \frac{(n^{2}-1)^{2}}{ n^{2} (n^{2}+1)} ; \\
a_{3} = \alpha^{2} \frac{ (n^{2}-1)^{2}}{2 n^{2} (n^{2}+1)}, \qquad a_{4} = \frac{\alpha^{4}}{2 n^{2} (n^{2}+1)}   
\end{split}
\end{equation}

The estimation of the different terms of equation \eqref{eq:nonlin} shows that the essential contribution get the first and the second nonlinear terms only.  In further we skip the last nonlinear terms. 

Eq. \eqref{eq:nonlin} allows us to calculate the eigenfrequencies in the linear approximation as well as to estimate the effect of nonlinearity on these frequencies at different boundary conditions.  
The detail analysis shows that  two first nonlinear terms in the equation \eqref{eq:nonlin} give the dominant contribution in the low-frequency dynamics of CNTs.
One should note, that the parametrs $\alpha$ and $\beta$ are small enough in most cases. 
Therefore, the parameter $\kappa$ is very close to the $\gamma$.
Moreover, the own frequency of the gap ($\omega_{0}$ is small due to the smalness of the effective thickness of the CNT wall). 
In such a case it is convenient to introduce the 'new' time scaled by the gap frequency $\omega_{0}$: $\tau_{0}=\omega_{0} \tau$.
Taking into account that the coefficients $\mu, \gamma, a_{1}, a_{2}$ are of the order of unity, one can rewrite the equation \eqref{eq:nonlin} keeping the dominant terms with the order of small parameters, which do not exceed two:

\begin{equation}\label{eq:nonlin2}
\begin{split}
 \frac {\partial^{2}W}{\partial \tau_{0}^{2}}+   W -\frac{\mu}{\omega_{0}^{2}} \frac{\partial^{2} W}{\partial \xi^{2}} -  \gamma\frac{\partial ^{4} W}{\partial \xi^{2} \partial \tau_{0}^{2}}+\frac{\kappa}{\omega_{0}^{2}}  \frac{\partial^{4} W}{\partial \xi^{4}}   \\
 + a_{1} W \left( \left( \frac{\partial W}{\partial \tau_{0}} \right) ^{2}+W \frac{\partial ^{2} W}{\partial \tau_{0}^{2}} \right) -   \frac{a_{2} }{\omega_{0}^{2}} \left( \frac{\partial W}{\partial \xi} \right)^{2} \frac{\partial ^{2} W}{\partial \xi^{2}} = 0,
\end{split}
\end{equation}
\section[level(1)]{The multiscale expansion}\label{app:A1}
Because we consider the small-amplitude oscillations, one can represent the complex amplitude $\Psi$ as a series of small parameter $\varepsilon$:

\begin{equation}\label{eq:series1}
\Psi = \varepsilon \left( \psi_{0} + \varepsilon \psi_{1} + \varepsilon^{2} \psi_{2} + \dots \right)
\end{equation}

Next we should introduce the 'time' series: $\tau_{1}=\varepsilon \tau_{0}, \tau_{2}=\varepsilon^2 \tau_{0}, \dots$ and the respective time derivatives: 

\begin{equation}\label{eq:timeexp}
 \frac{\partial}{\partial \tau_{0}}=\frac{\partial}{\partial \tau_{0}}+\varepsilon \frac{\partial}{\partial \tau_{1}} + \varepsilon^{2} \frac{\partial}{\partial \tau_{2}} + \dots  
 \end{equation}

Substituting the expansion \eqref{eq:series1} into \eqref{eq:WWW} and taking into account the hierarchy of the times, we get the equations in the different orders by small parameter $\varepsilon$.

\begin{equation*}\label{eq:order1}
\varepsilon^{1}: \quad \quad
i \frac{\partial \psi_{0}}{\partial \tau_{0}}-\psi_{0} = 0
\end{equation*}

So, we get

$$ \psi_{0}=\chi_{0}(\xi, \tau_{1}, \tau_{2}) e^{- i \tau_{0}} $$

\begin{equation*}\label{eq:order2}
\begin{split}
\varepsilon^{2}: \quad \quad
 i  \frac{\partial \psi _{1}}{\partial \tau_{0}}-\frac{\partial \psi _{0}}{\partial \tau_{1}} - \psi _{1}=0
\end{split}
\end{equation*}

Then we get:

\begin{equation}\label{eq:psidef2}
\begin{split}
\psi_{1}(\xi, \tau_{0},\tau_{1},\tau_{2})= \chi_{1}(\xi, \tau_{1}, \tau_{2}) e^{-i \tau_{0}}   \\  
\psi_{0}(\xi,\tau_{0},\tau_{1},\tau_{2}) = \chi_{0}(\xi,\tau_{2}) e^{-i \tau_{0}} .
\end{split}
\end{equation}


\begin{multline}\label{eq:order3}
\varepsilon^{3} :  \\
 i \frac{\partial \psi_{2}}{\partial \tau_{0}} -  \psi_{2} + i \frac{\partial \psi_{1}}{\partial \tau_{1}} + i \frac{\partial \psi_{0}}{\partial \tau_{2}}  - \frac{\mu}{2 \omega_{0}^{2}} \frac{\partial^{2} \psi_{0}}{\partial \xi^{2}}  \\  -\frac{\gamma}{2}\frac{\partial^{2} \psi_{0}}{\partial \xi^{2}}  - i \frac{\gamma}{2} \frac{\partial^{3} \psi_{0}}{\partial \tau_{0} \partial \xi^{2}}  +\frac{\kappa} {2 \omega_{0}^{2} }\frac{\partial^{4} \psi_{0}}{\partial \xi^{4}}   \\  + \frac{\mu}{2 \omega_{0}^{2}} \frac{\partial^{2} \psi_{0}^{*}}{\partial \xi^{2}} - \frac{\gamma}{2}\frac{\partial^{2} \psi_{0}^{*}}{\partial \xi^{2}}   - \frac{\kappa} {2 \omega_{0}^{2} }\frac{\partial^{4} \psi_{0}^{*}}{\partial \xi^{4}}   \\
+\frac{a_{1} }{2} \left( | \psi_{0} |^{2}  \psi_{0} -i \frac{\partial \psi_{0}}{\partial \tau_{0}} \left( \psi_{0}^{2} + \psi_{0}^{*2}  - 2 |\psi_{0} |^{2} \right) \right)= 0,
\end{multline}

Taking into account the relations \eqref{eq:psidef2}, one can integrate the equation \eqref{eq:order3} with respect to $\tau_{0}$ and $\tau_{1}$. 
Then the condition of the secular terms absence give us the equation for the main approximation amplitude $\chi_{0}$ \eqref{eq:NLSE}.

 
 \section[level(1)]{Influence of boundary conditions}\label{app:B}

The presence of boundary conditions different from the simple supporting affects on the NNM and their frequency. 
In this Appendix we consider the effective method for the solution of boundary problem and demonstrate the procedure of the normal mode construction on the example of CNT with free edges.

It is intuitively clear that the strong boundary conditions like the clamping lead to frequency growth while the more ''soft'' conditions can decrease the frequencies. To estimate the variation of normal modes we used the linear approximation of RSKTST (reduced Sanders Koiter thin shell theory). 

Let us assume that the solution of linearized equation of the the CNT vibrations 

\begin{equation}\label{eq:lin}
 \frac{\partial^{2}W}{\partial \tau^{2}}+  \omega_{0}^{2} W -\mu \frac{\partial^{2} W}{\partial \xi^{2}} -  \gamma \frac{\partial ^{4} W}{\partial \xi^{2} \partial \tau^{2}}+\kappa \frac{\partial^{4} W}{\partial \xi^{4}}   = 0
 \end{equation}

is  represented as the periodic process

\begin{equation}\label{eq:W}
W(\xi, \tau) \sim f(\xi) \cos(\omega \tau).
\end{equation} 

Taking into account expression \eqref{eq:W} explicitly, one can rewrite equation \eqref{eq:lin}  with the help of the product of two differential operators:

\begin{equation}
\kappa \left(\frac{d^{2}}{d \xi^{2}} + k^{2}\right) \left(\frac{d^{2}}{d \xi^{2}} - \lambda^{2}\right) f = 0,
\label{LS:factordiff}
\end{equation}

where the parameters $\mu$,  $\gamma$, and $\omega$ are linked by the relationships
\begin{equation}
\begin{split}
\kappa \lambda ^2 k^2=\omega ^2-\omega_{0} ^2 \\
\kappa \left(\lambda ^2-k ^2\right)=\mu-\gamma \omega ^2. \\
\label{eq:edge1}
\end{split}
\end{equation}

Because the operators
 $ \left( d^{2} / d \xi^{2} + k^{2} \right) $
 and 
$ \left( d^{2} / d \xi^{2} - \lambda^{2} \right) $
  are commutative ones, any function $f( \xi)$, which satisfies one of the equations
 \begin{equation}\label{eq:B7}
 \left(  \frac{d^{2}}{d \xi^{2}} + k^{2} \right) f( \xi)=0
 \end{equation}
 
 \begin{equation}\label{eq:B8}
 \left(   \frac{d^{2}}{d \xi^{2}} - \lambda^{2} \right) f( \xi)=0,
 \end{equation}
 is a solution of equation \eqref{LS:factordiff}.

 So a general solution of equation \eqref{LS:factordiff} is a linear combination of the solutions of equations \eqref{eq:B7}, \eqref{eq:B8}:
\begin{equation}\label{eq:genform}
\begin{split}
 f( \xi)=( C_{1} \sin{ k (\xi- \xi_{0})}+ \\ C_{2} e^{ \lambda (\xi - \xi_{1})} +C_{3} e^{- \lambda ( \xi - \xi_{1})} ) ,
\end{split}
\end{equation}

where $k$, $\lambda$, $C_{j}, j=1,2,3$ and $\xi_{j}, j=0,1$ are the constants determined by the boundary conditions and the symmetry of solution. 
Equation \eqref{eq:genform}  shows that the proposed approach allows clearly single out the exponential boundary layer as a part of the solution. 
One should note that expressions \eqref{eq:edge1} provide the coupling between the parameters of the solution.
The estimation of the parameters of solution \eqref{eq:genform} is needed for formulation of the boundary conditions in terms of the radial displacements. 

Let us consider the vibrations of CNT under condition of free edges. 
One can show that the free edges boundary conditions correspond to two equations for the radial displacement $W(\xi, \tau)$:

\begin{equation}\label{eq:freeedgeW}
\left.
\begin{aligned}
\alpha ^2 \frac{\partial ^{2} W(\xi, \tau )}{\partial \xi^{2}}-\nu  \left( n^{2}-1 \right) W(\xi, \tau )=0 \\
\alpha ^2 \frac{\partial ^{3} W(\xi, \tau )}{\partial \xi^{3} }-(2-\nu ) \left(n^2-1\right) \frac{\partial W(\xi, \tau )}{\partial \xi }=0
\end{aligned}
\right\} 
\quad \xi=0, 1
\end{equation}

Solving the equations (\ref{eq:edge1}-\ref{eq:freeedgeW}, one can estimate all the parameters of the solution \eqref{eq:genform}.

\begin{figure}[!ht]
\center{

\includegraphics[width = 70mm, height = 45mm]{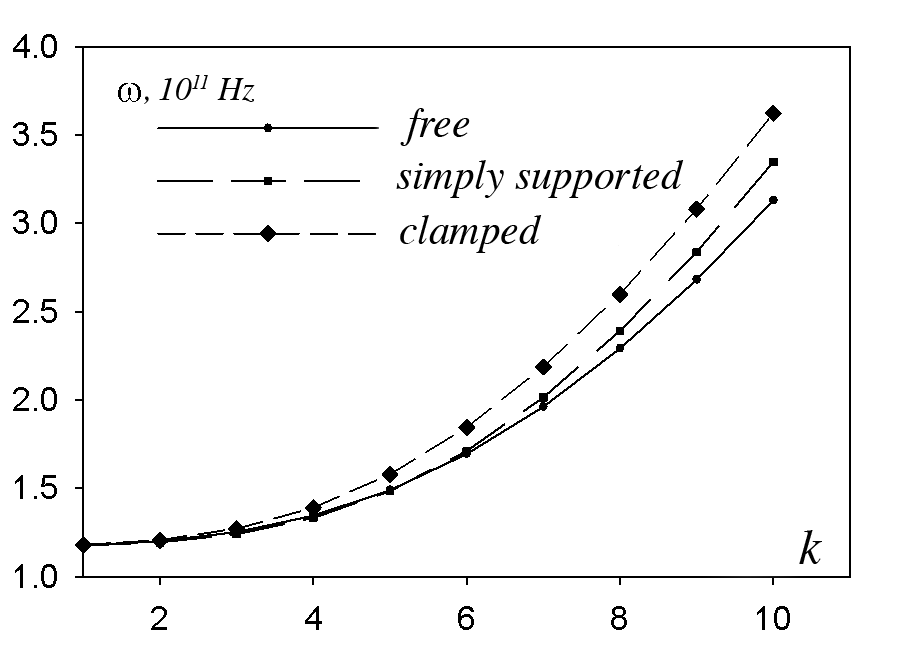}

}
\caption{The comparison of CNT vibration spectra for the different boundary conditions. CNT parameters -  $L=10$ nm  and $R = 0.79$ nm.}
\label{LS:ris:spectra}
\end{figure}

Fig. \ref{LS:ris:spectra} shows the spectra for the system under different boundary conditions - the simply supported, free and clamped edges.

%
%

\bibliography{Beat_Loc}

\end{document}